\documentclass[twocolumn]{aastex631}
\usepackage[T1]{fontenc}
\usepackage[utf8]{inputenc}
\let\tablenum\relax
\usepackage{mathtools}
\usepackage{siunitx}
\usepackage{physics}
\usepackage{graphicx}
\usepackage{enumerate}
\usepackage{url}
\usepackage{comment}

\newcommand{\Zsun}{{\si{Z_\odot}}}
\newcommand{\Msun}{{\si{M_\odot}}}
\newcommand{\Mh}{{M_\mathrm{h}}}
\newcommand{\Macc}{{\dot{M}_\mathrm{h}}}
\newcommand{\Mmol}{{M_\mathrm{mol}}}
\newcommand{\Matom}{{M_\mathrm{atom}}}

\newcommand{\etaIII}{{\eta_\mathrm{III}}}
\newcommand{\SFR}{{\dot{M}_\star^\mathrm{III}}}
\newcommand{\UVLF}{\Phi_\mathrm{UV}^\mathrm{III}}
\newcommand{\MUV}{M_\mathrm{UV}^\mathrm{III}}
\newcommand{\LUV}{L_\mathrm{UV}^\mathrm{III}}
\newcommand{\kappaUV}{\kappa_\mathrm{UV}^\mathrm{III}}
\newcommand{\epsilonstar}{{\varepsilon_\star^\mathrm{III}}}
\newcommand{\epsilonstarUV}{{\varepsilon_\mathrm{\star,UV}^\mathrm{III}}}
\newcommand{\betastar}{{\beta_\star^\mathrm{III}}}
\newcommand{\alphastar}{{\alpha_\star^\mathrm{III}}}
\newcommand{\Mup}{{M_\mathrm{up}^\mathrm{III}}}
\newcommand{\Mp}{{M_\mathrm{p}^\mathrm{III}}}
\newcommand{\sigmaUV}{{\sigma_\mathrm{UV}^\mathrm{III}}}

\graphicspath{{./}{figures_v1/}}
\begin{document}

\title{Bursty or heavy? The surprise of bright Population~III systems in the Reionization era}

\correspondingauthor{Alessandra Venditti}
\email{alessandra.venditti@utexas.edu}

\author[0000-0003-2237-0777]{Alessandra Venditti}
\affiliation{Department of Astronomy, University of Texas at Austin, 2515 Speedway, Stop C1400, Austin, TX 78712, USA}
\affiliation{Cosmic Frontier Center, The University of Texas at Austin, Austin, TX 78712}
\begingroup
\renewcommand\thefootnote{}\footnotetext{* Cosmic Frontier Center Prize Fellow}
\endgroup

\author[0000-0002-8984-0465]{Julian B. Muñoz}
\affiliation{Department of Astronomy, University of Texas at Austin, 2515 Speedway, Stop C1400, Austin, TX 78712, USA}
\affiliation{Cosmic Frontier Center, The University of Texas at Austin, Austin, TX 78712}

\author[0000-0003-0212-2979]{Volker Bromm}
\affiliation{Department of Astronomy, University of Texas at Austin, 2515 Speedway, Stop C1400, Austin, TX 78712, USA}
\affiliation{Cosmic Frontier Center, The University of Texas at Austin, Austin, TX 78712}
\affiliation{Weinberg Institute for Theoretical Physics, University of Texas at Austin, Austin, TX 78712, USA}

\author[0000-0001-7201-5066]{Seiji Fujimoto$^{*}$}
\affiliation{David A. Dunlap Department of Astronomy and Astrophysics, University of Toronto, 50 St. George, Toronto, ON M5S 3H4, Canada}
\affiliation{Dunlap Institute for Astronomy and Astrophysics, 50 St. George, Toronto, ON M5S 3H4, Canada}
\affiliation{Department of Astronomy, University of Texas at Austin, 2515 Speedway, Stop C1400, Austin, TX 78712, USA}
\begingroup
\renewcommand\thefootnote{}\footnotetext{* Hubble Fellow at the University of Texas at Austin}
\endgroup

\author[0000-0001-8519-1130]{Steven L. Finkelstein}
\affiliation{Department of Astronomy, University of Texas at Austin, 2515 Speedway, Stop C1400, Austin, TX 78712, USA}
\affiliation{Cosmic Frontier Center, The University of Texas at Austin, Austin, TX 78712}

\author[0000-0002-0302-2577]{John Chisholm}
\affiliation{Department of Astronomy, University of Texas at Austin, 2515 Speedway, Stop C1400, Austin, TX 78712, USA}
\affiliation{Cosmic Frontier Center, The University of Texas at Austin, Austin, TX 78712}

%% Note that the \and command from previous versions of AASTeX is now
%% depreciated in this version as it is no longer necessary. AASTeX 
%% automatically takes care of all commas and "and"s between authors names.

%% AASTeX 6.31 has the new \collaboration and \nocollaboration commands to
%% provide the collaboration status of a group of authors. These commands 
%% can be used either before or after the list of corresponding authors. The
%% argument for \collaboration is the collaboration identifier. Authors are
%% encouraged to surround collaboration identifiers with ()s. The 
%% \nocollaboration command takes no argument and exists to indicate that
%% the nearby authors are not part of surrounding collaborations.

%% Mark off the abstract in the ``abstract'' environment. 
\begin{abstract}
The nature of the first, so-called Population~III (Pop~III) stars has for long remained largely unconstrained. However, the James Webb Space Telescope (JWST) finally opened new concrete prospects for their detection during the Epoch of Reionization (EoR), notably providing promising observational constraints on the Pop~III ultra-violet luminosity function (UVLF) at $z \approx 5.6 - 6.6$. These preliminary data hint towards an unexpected population of UV-bright Pop~III sources, which challenges the prevailing view that Pop~III star formation is confined to molecular-cooling mini-halos. 
Here we show that there are two families of models that can explain these surprising observations, either by allowing for late-time Pop~III formation within massive, atomic-cooling halos (with halo masses up to $\Mup \gtrsim 10^{10} ~\Msun$) or by invoking a highly bursty Pop~III star-formation activity (with a stochasticity parameter $\sigmaUV \gtrsim 1.5$). In these scenarios, Pop~III systems would have to be either heavier or burstier than usually assumed, underscoring the need to reconsider common assumptions about Pop~III star-formation sites, and the potential implications of JWST candidates for current and future observations.
\end{abstract}

%% Keywords should appear after the \end{abstract} command. 
%% The AAS Journals now uses Unified Astronomy Thesaurus concepts:
%% https://astrothesaurus.org
%% You will be asked to selected these concepts during the submission process
%% but this old "keyword" functionality is maintained in case authors want
%% to include these concepts in their preprints.
 % to be updated following APJ restrictions

%% From the front matter, we move on to the body of the paper.
%% Sections are demarcated by \section and \subsection, respectively.
%% Observe the use of the LaTeX \label
%% command after the \subsection to give a symbolic KEY to the
%% subsection for cross-referencing in a \ref command.
%% You can use LaTeX's \ref and \label commands to keep track of
%% cross-references to sections, equations, tables, and figures.
%% That way, if you change the order of any elements, LaTeX will
%% automatically renumber them.
%%
%% We recommend that authors also use the natbib \citep
%% and \citet commands to identify citations.  The citations are
%% tied to the reference list via symbolic KEYs. The KEY corresponds
%% to the KEY in the \bibitem in the reference list below. 

\section{Introduction} 
\label{sec:intro}

The traditional theoretical framework for Pop~III star formation predicts that these stars primarily form from primordial gas in molecular-cooling mini-halos, with masses $\Mh \sim 10^5 - 10^6 ~\Msun$ at Cosmic Dawn ($z \sim 20 - 30$, assuming no/weak Lyman-Werner or LW feedback; e.g., reviewed in \citealt{Bromm_2013, Klessen_Glover_2023}). The dominant coolant in these environments (H$_2$) is significantly less efficient than metal-line cooling, which is prevalent in typical enriched regions. As a result, Pop~III stars are expected to form with much lower star-formation efficiencies (SFEs) compared to present-day stars, exhibiting a top-heavy initial mass function (IMF), possibly extending up to $\sim 10 - 10^{2} ~\Msun$ \citep{Hosokawa_2011, Susa_2014, Stacy_2016, Chon_2024, Mana_Omukai_2024, Tang_Chen_2024} or even $\sim 10^{3} ~\Msun$ \citep{Hirano_2014, Hirano_2015, Hosokawa_2016, Sugimura_2020, Latif_2022}. However, both theoretical models and emerging observations suggest that this picture may be incomplete.

Cosmological simulations \citep{Johnson_2013, Xu_2016_X-ray, Xu_2016, Jaacks_2019, Skinner_Wise_2020, Zier_2025} and semi-analytical models \citep{Magg_2018, Mebane_2018, Visbal_2020, Trinca_2024, Ventura_2024, Ventura_2025,Liu_2025_SAM} indicate that Pop~III star formation can persist well after its onset at Cosmic Dawn, down to the EoR ($z \sim 6 - 10$). Moreover, Pop~III stars at this epoch can form in halos much larger than the first mini-halos \citep{Bennett_Sijacki_2020, Liu_Bromm_2020, Riaz_2022, Venditti_2023}. In fact, pristine, star-forming gas pockets could survive even within globally enriched halos that already experienced star-forming episodes, thanks to the in-homogeneous and patchy nature of metal enrichment; this phenomenon is further enhanced in models including incomplete sub-grid mixing \citep[e.g.,][]{Ji_Mixing2015,Sarmento_2018, Sarmento_Scannapieco_2022}. Another possibility is that star formation is delayed by the presence of a sufficiently strong LW radiation field from previous generations of stars: in externally irradiated halos, H$_2$ cooling is suppressed until they become dense enough to shield themselves from the radiation \citep[e.g.,][]{Wolcott-Green_2019}. We note that such ``LW-bubbles'' propagate into the early intergalactic medium (IGM) much faster than the corresponding ``metal-bubbles''; both originate in the same stars, but the former effectively expand at the speed of light, whereas the latter at close to the IGM sound speed\footnote{The physical distance traveled through the neutral IGM by a LW photon until it is redshifted to the closest hydrogen Lyman-series transition, and thus absorbed, is of order $r_\mathrm{max} \sim 100$~cMpc \citep[e.g.,][]{Ahn_2009, Haiman_2000,Johnson_2008, Ahn_2009}.}. Once the halos have accreted enough mass, they reach virial temperatures of $\sim 8000$~K and enter the so-called ``atomic''- or ``Ly$\alpha$''-cooling regime, finally allowing star formation to occur \citep{Oh_Haiman_2002, Agarwal_2019}. Early investigations by \citet{Greif_Bromm_2006, Greif_2008} indicate that primordial gas collapsing into atomic-cooling halos at $z \lesssim 10$ might experience a boost in ionization (e.g., through shocks), resulting in more efficient HD cooling and a consequent boost in SFE compared to mini-halos at Cosmic Dawn \citep{Johnson_Bromm_2006}. This mode of star formation has previously been referred to as ``Pop~III.2'', to differentiate it from the standard ``Pop~III.1'' formation pathway associated with H$_2$ cooling in mini-halos \citep{Bromm_2009}. 

Recent JWST observations are shedding new light into Pop~III star formation through the study of the galaxy UVLF. A top-heavy IMF component, typical of Pop~III stars \citep{Inayoshi_2022, Finkelstein_2023, Harikane_2023, Harikane_2024, Yung_2024, Trinca_2024, Ventura_2024, Cueto_2024, Hutter_2025, Lu_2025, Harvey_2025, Jeong_2025, Mauerhofer_2025}, has been suggested to explain the surprising overabundance of UV-bright galaxies at $z \gtrsim 10$ \citep{Castellano_2022, Naidu_2022, Finkelstein_2023, Finkelstein_2024, Bouwens_2023, Bouwens_2023_UVLF, Harikane_2023, Harikane_2024, Perez-Gonzalez_2023, Robertson_2024, McLeod_2024, Adams_2024, Kokorev_2025}\footnote{A variety of other solutions have been proposed to solve this apparent tension, including burstier star-formation histories \citep[e.g.,][]{Mason_2023}, feedback-free bursts \citep[e.g.,][]{Dekel_2023}, density-modulated star-formation models \citep{Somerville_2025}, and attenuation-free models \citep[e.g.,][]{Ferrara_2023}.}. 
However, only one potential Pop~III-hosting system has been spectroscopically identified at such high redshifts, i.e., the candidate HeII emitter in GN-z11 observed by \citet{Maiolino_2024} at $z \approx 10.6$. On the other hand, HeII emitters possibly associated with Pop~III stars have been discovered at lower redshifts (particularly, from \citealt{Wang_2024}  $z \approx 8.1$, and from \citealt{Vanzella_2020, Vanzella_2023} at $z \approx 6.7$). \citet{Schauer_2022} further attributed a non-negligible probability to a Pop~III origin of the lensed ``Earendel'' source within the Sunrise Arc at $z \approx 6.2$ \citep{Welch_2022}. Finally, an absence of metal lines has been reported for the AMORE6 galaxy at $z \approx 5.7$ \citep{Morishita_2025}, and for the MPG-CR3 galaxy at redshift as low as $z \approx 3.2$ \citep{Cai_2025}.

Recently, \citet{Fujimoto_2025} provided tantalizing preliminary constraints on the Pop~III UVLF, by applying a NIRCam-based selection criterion on galaxies across five JWST legacy fields (spanning a total area of $\approx 500$~arcmin$^2$) to capture Pop~III candidates at $z \approx 5.6 - 6.6$. A promising Pop~III candidate (GLIMPSE-16043, with UV magnitude $\approx -15.9$) was discovered at $z \approx 6.5$ within the GLIMPSE\footnote{\url{http://www.jwst-glimpse.com/}} \citep{Atek_2023, Kokorev_2025} field, one of the deepest (lensed) NIRCam fields so far. Photometry on this object initially exhibited key Pop~III features -- e.g., strong H$\alpha$ lines and Balmer jump, no dust, and undetectable metal lines -- which hinted towards the presence of a young ($< 5$\,Myr), metal-poor ($< 0.005 ~\Zsun$) stellar population of $\sim 10^5\,\Msun$, significantly deviating from the observed UV-metallicity relation at $z \sim 4 - 10$ \citep{Nakajima_2023, Chemerynska_2024}.
However, recent follow-up spectroscopy has detected a metal line and unveiled that unusual nebular conditions can mimic the Pop~III-like SED properties, raising an additional cautionary concern when using the photometric selection scheme (S.~Fujimoto et al. in prep; private communication)\footnote{Also note that, while the photometric-selection criterion relies on the non-detection of the bright [OIII] line, other mechanisms may contribute to a decreased brightness of this line, such as collisional de-excitation in dense environments \cite[see e.g.][]{Taylor_2024}.}. Another tentative candidate (JOF-21739, with an even brighter UV magnitude of $\approx -17.6$) has further been identified in the JADES Origin Field (JOF, \citealt{Robertson_2024}) at $z \approx 6.2$ -- only marginally consistent with the selection criteria.

The claimed Pop~III systems AMORE6, GLIMPSE-16043 and JOF-21739 would be at odds with previous theoretical expectations, where attention had been predominantly focused on faint formation sites at cosmic dawn ($z\gtrsim 15$), further assuming that afterwards metal-enriched (Pop~II/I) star formation would completely dominate (including at the EoR).
Motivated by these new observations, in the present work we explore the role of two late Pop~III-formation scenarios in shaping the bright end of the Pop~III UVLF at late times: (i) an efficient Pop~III star-formation component extended into massive atomic-cooling halos (``\textsc{Heavy}'' mode); (ii) a bursty Pop~III star-formation activity confined within molecular-cooling mini-halos (``\textsc{Bursty}'' mode).

We summarize the model in Section~\ref{sec:methodology} and present fits for the two models at $z \approx 5.6 - 6.6$ in Section~\ref{sec:z6.1_fit}. We discuss implications and perspectives for future observations at deeper UV magnitudes and higher redshifts in Section~\ref{sec:deepMUV_high-z}, and conclude in Section~\ref{sec:conclusions}.
Throughout this paper we use AB magnitudes~\citep{Oke_1974} and a flat~\citet{Planck_2020} cosmology.

\section{Modeling the Pop~III UVLF} 
\label{sec:methodology}

We will focus on the Pop~III UVLF ($\UVLF$), which measures the number density of Pop~III systems with a given UV magnitude $\MUV$ as:
\begin{equation}
    \UVLF \equiv \dv{n}{\MUV} = \int \dv{n}{\Mh} \dd{\Mh} P(\MUV | \Mh),
    \label{eq:UVLF_III}
\end{equation}
where $\dd{n}/\dd{\Mh}$ is the halo mass function (modeled through the public cosmology python package \texttt{classy}\footnote{\url{https://cobaya.readthedocs.io/en/latest/theory_class.html}}, \citealt{Blas_2011_classy}, following \citealt{Sheth_Tormen_2002, Rodriguez-Puebla_2016}), and $P(\MUV | \Mh)$ is the conditional probability that a halo of mass $\Mh$ hosts a Pop~III system with magnitude $\MUV$. The latter term implicitly encodes the halo-galaxy connection, and it is modeled as a Gaussian with a mass-independent dispersion $\sigmaUV$, centered around the mean value given by the UV luminosity
\begin{equation}
    \LUV (\Mh) = \SFR (\Mh) / \kappaUV,
    \label{eq:LUV_III}
\end{equation}
where $\kappaUV$ is the conversion factor between the Pop~III star-formation rate (SFR, $\SFR$) and the Pop~III UV luminosity\footnote{The UV magnitude corresponding to a given specific UV luminosity is given by:
$
    M_\mathrm{UV} \simeq 51.63 - 2.5 \, \mathrm{log} \left( \frac{L_\mathrm{UV}}{\si{erg.s^{-1}.Hz^{-1}}} \right).
$
} ($\LUV$). The UVLF is computed in bins of $\MUV$ and redshift ($z$), assuming top-hat/Gaussian window functions around the central $\MUV$/$z$ values respectively.

We adopt the model of \citet[][see also~\citealt{Munoz_2023_Zeus21, Qin_2020}]{Cruz_2024} for the average Pop~III SFR of systems hosted in halos of mass $\Mh$:
\begin{equation}
    \SFR (\Mh) = f_\star^\mathrm{III} (\Mh) f_\mathrm{duty}^\mathrm{III} (\Mh) f_\mathrm{b} \Macc,
    \label{eq:SFR_III}
\end{equation}
where $f_\mathrm{b} \approx 0.16$ is the cosmic baryon fraction \citep{Planck_2020}, and assume a double power-law functional form for the star-formation efficiency (SFE)
\begin{equation}
    f_\star^\mathrm{III} (\Mh) = \frac{2\epsilonstar}{(\Mh/\Mp)^\alphastar + (\Mh/\Mp)^\betastar},
    \label{eq:fstar_III}
\end{equation}
which encodes feedback as in the case of Pop~II galaxies~\citep{Moster_2013, Furlanetto_2017, Sabti_2022}.  Similar to \citet{Cruz_2024}, we will assume a flat SFE with $\alphastar = \betastar = 0$, corresponding to a low-feedback scenario, and $\Mp \sim 10^7 ~\Msun$, while the amplitude $\epsilonstar$ is free to vary. 
We also incorporate a duty-cycle
\begin{equation}f_\mathrm{duty}^\mathrm{III} (\Mh) = \exp(-\Mmol/\Mh) \exp(-\Mh/\Mup),
    \label{eq:fduty_III}
\end{equation}
that parametrizes the fraction of halos hosting Pop~III stars as a function of $\Mh$. 
It is often assumed that Pop~III galaxies form above the molecular-cooling limit, until a threshold mass $\Mup = \Matom (z) \simeq 3.3 \times 10^7 ~\Msun [(1 + z)/21]^{3/2}$ (corresponding to a virial temperature of $\sim 10^4$~K, \citealt{Oh_Haiman_2002}), as systems begin to retain metals and transition from Pop~III to Pop~II star formation.
Here, instead, we relax this assumption and allow the upper mass cut-off $\Mup$ to vary as a free parameter of the model.
Note that the exact value of $\Mmol(z)$ (as well as $\alphastar$) are of limited interest for the present work, as they predominantly affect the faint end of the UVLF, beyond direct constraints (but see the discussion in Appendix~\ref{app:further_modeling}). Our choice of $\betastar = 0$ is also examined in Appendix~\ref{app:further_modeling}, where we study the constraints on $\betastar$ imposed by observations with fixed values of $\Mup$ -- strongly degenerate with $\betastar$.

The conversion factor $\kappaUV$ in Equation~\ref{eq:LUV_III} is also fully degenerate with the normalization $\epsilonstar$ in Equation~\ref{eq:fstar_III}. Therefore, we can define an effective UV-emission efficiency~\citep{Munoz_2023}
\begin{equation}
    \epsilonstarUV \equiv \epsilonstar (\kappaUV / \kappa_\mathrm{UV,ref})^{-1},
    \label{eq:epsilonstar_UV}
\end{equation}
where $\kappa_\mathrm{UV,ref} \approx 1.15 \times 10^{-28} ~(\si{M_\odot\,yr^{-1}})/(\si{erg.s^{-1}})$ is the fiducial value for standard Pop~II stellar populations from \citet[][not including dust attenuation]{Madau_Dickinson_2014}. We note that \citet{Inayoshi_2022} suggested up to $\sim 3-4$ times higher UV luminosities for extremely top-heavy IMFs, more typical of Pop~III systems, while even higher values (up to a factor six) have been discussed in the case of rapidly rotating stars undergoing chemically homogeneous evolution \citep{Liu_2025}.
With this parametrization, the mean $\epsilonstarUV$ and the stochasticity $\sigmaUV$ encompass variations in the average value and scatter of both the underlying star-formation activity in halos and their UV luminosity per unit SFR. This formalism takes into account a potential enhanced UV luminosity per unit SFR with respect to standard Pop~II stellar populations, and a potential increased burstiness in the star-formation activity. In the reference case, we consider $\sigmaUV = 0.7$, as in the best fit at $z \approx 5.6 - 6.6$ of \citet{Munoz_2023} for Pop~II (calibrated against the observed UVLF at $z \sim 4 - 8$ from \citealt{Bouwens_2021}), but we explore enhanced burstiness in Section~\ref{sec:z6.1_fit_bursty} and in Appendix~\ref{app:further_modeling}.

\section{The redshift $\approx 5.6 - 6.6$ Pop~III UVLF}
\label{sec:z6.1_fit}

\subsection{Pop~III star formation beyond mini-halos}
\label{sec:z6.1_fit_heavy}

\begin{figure*}
    \centering
    \includegraphics[width=0.49\linewidth]{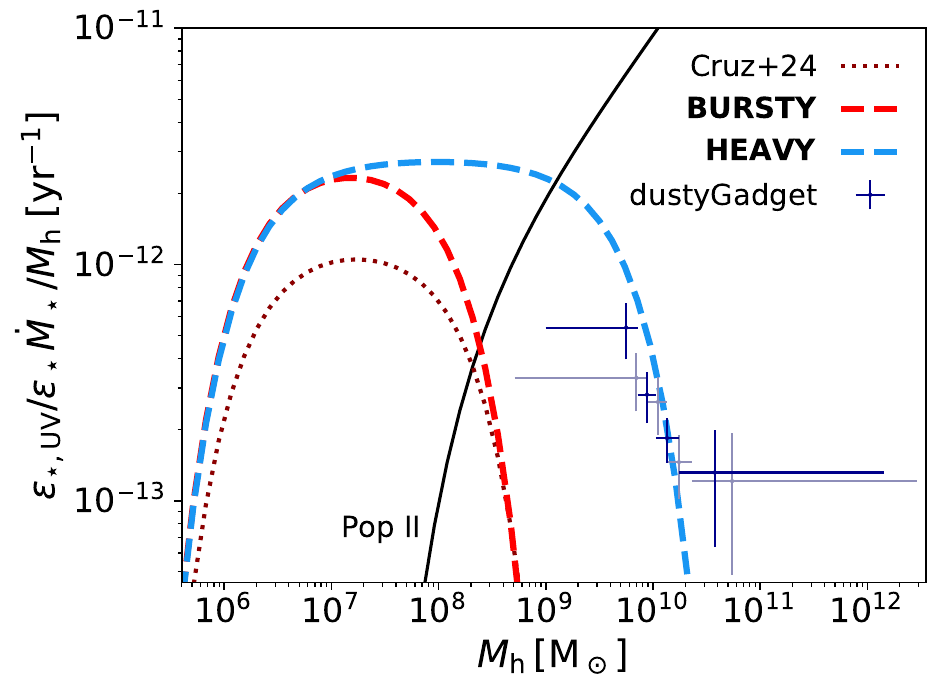}
    \includegraphics[width=0.49\linewidth]{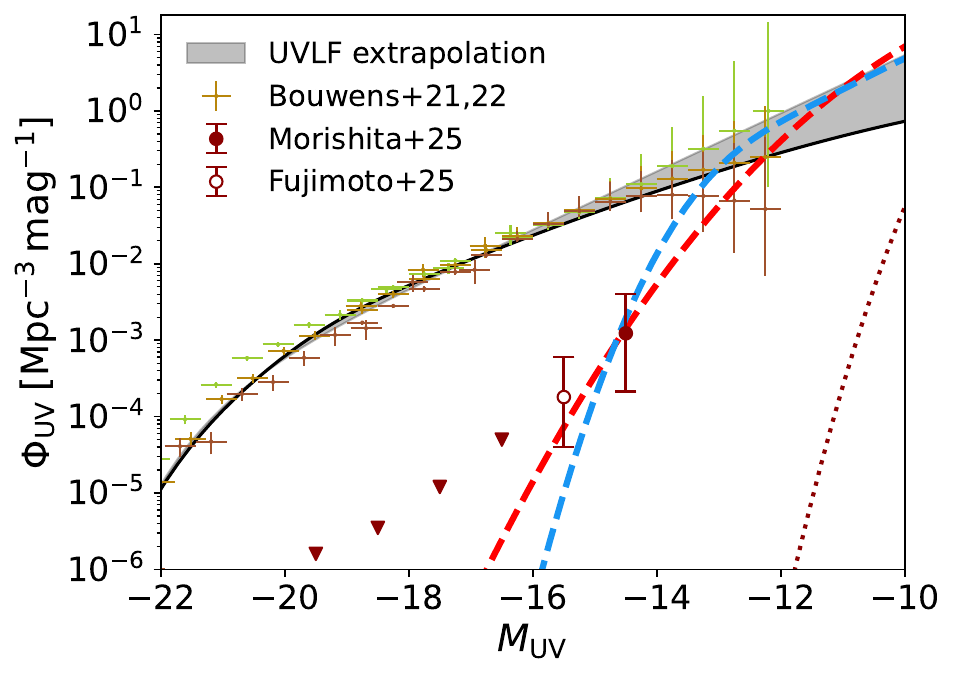}
    \caption{
    \textbf{Left:} SFR per unit halo mass ($\dot{M}_\star / \Mh$) as a function of halo mass ($\Mh$) at $z = 6.1$ (with $\Delta z = 1)$, incorporating a UV enhancement of $\varepsilon_\mathrm{\star,UV} / \varepsilon_\star$. Our reference ``\textsc{Bursty}''/``\textsc{Heavy}'' models (with $\Mup \sim 10^8/10^{10} ~\Msun$ respectively, and $\epsilonstarUV \sim 10^{-2.4}$) are shown as \textit{red/blue, thick, dashed lines}. The model of \citet[][\textit{dark-red, dotted line}, see text for details]{Cruz_2024}, and the corresponding Pop~II model from \citet[][\textit{black, solid line}]{Munoz_2023}, are included as a reference. The \textit{blue crosses} show the average $\dot{M}_\star / \Mh$ at the two extremes of the considered redshift range ($z \approx 5.6/6.6$, in \textit{lighter/darker shade} respectively) for the \texttt{dustyGadget} simulations \citep{DiCesare_2023, Venditti_2023} in bins of $\Mh$, assuming a rescaling of $\etaIII = 0.3$ for the Pop~III mass (see Appendix~\ref{app:model_details} for more details). 
    %%%%%%%%%%%%%%%%%%%%%%%%%%%%%%%%%%%%%%%%%
    \textbf{Right:} UVLFs ($\Phi_\mathrm{UV}$) as a function of UV magnitude ($M_\mathrm{UV}$) resulting from the models in the left panel (with same color and linestyle, assuming $\sigmaUV = 1.5/0.7$ for the ``\textsc{Bursty}''/``\textsc{Heavy}'' models, respectively), compared with the data point and upper limits from \citet[][\textit{dark-red circle} and \textit{triangles}]{Fujimoto_2025}, including the AMORE6 candidate from \citet{Morishita_2025} in the faintest $M_\mathrm{UV}$ bin; \textit{filled/empty circles} distinguish between $M_\mathrm{UV}$ bins containing spectroscopic/photometric candidates (see text and Appendix~\ref{app:alternative_datasets} for more details). The \textit{grey, shaded area} encompasses a range of UVLF fitting functions (from \citealt{Bouwens_2021, Finkelstein_Bagley_2022, Munoz_2023}) extrapolated at the faint end, included as a reference together with data points from \citet{Bouwens_2021} and \citet{Bouwens_2022} at $z = 5/6/7$ (\textit{green/golden/brown dots}). While the model of \citet{Cruz_2024} assuming $\sigmaUV = 0.7$ (consistent with the value inferred from galaxy populations at the same redshift, as demonstrated by the agreement with observed data points of the Pop~II \textit{black line} when assuming $\sigmaUV \approx 0.7$) is unable to explain the data, models with enhanced stochasticity (``\textsc{Bursty}'' model) or a larger high-mass cut-off (``\textsc{Heavy}'' model) would be marginally consistent with GLIMPSE-16043 and AMORE6.}
    \label{fig:PopIII_UVLF_ref}
\end{figure*}

\begin{figure}
    \centering
    \includegraphics[width=\linewidth]{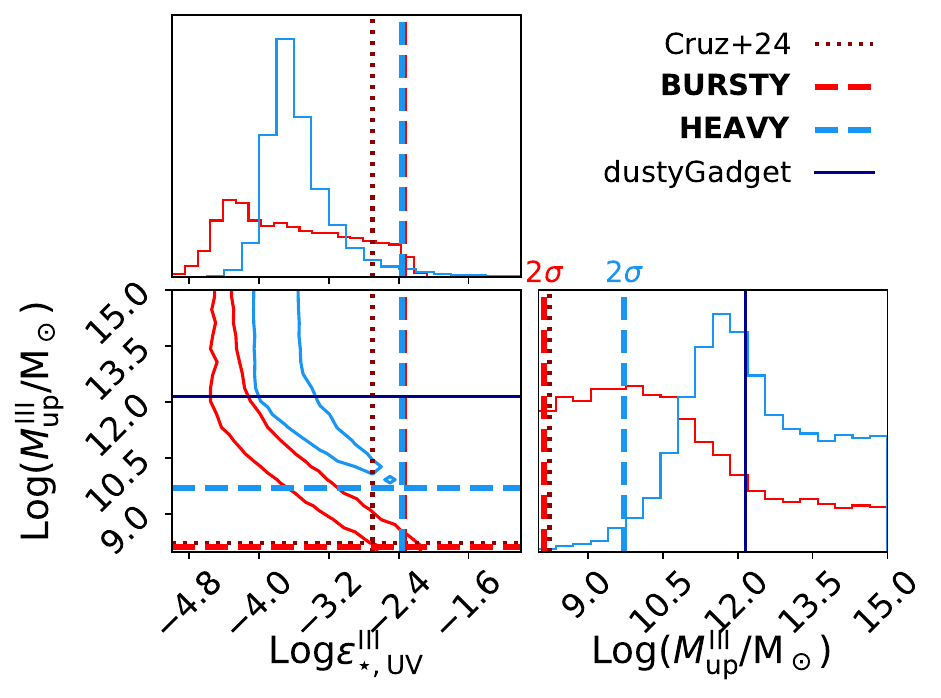}
    \caption{Joint posterior for the effective SFE normalization $\epsilonstarUV$ (Equation~\ref{eq:epsilonstar_UV}) and the high-mass cut-off $\Mup$ (Equation~\ref{eq:fduty_III}), resulting by fitting our Pop~III UVLF model described in Section~\ref{sec:methodology} against the observed data points from \citet{Fujimoto_2025} and \citet{Morishita_2025} at $z \approx 5.6 - 6.6$, with fixed $\betastar = 0$ and $\sigmaUV = 1.5/0.7$ (\textit{red/blue curves}). The \textit{dashed, thick lines} show the values adopted for $\epsilonstarUV$ and $\Mup$ in our reference ``\textsc{Bursty}'' and ``\textsc{Heavy}'' models (refer to Figure~\ref{fig:PopIII_UVLF_ref}): for each chosen value of $\sigmaUV$, these represent the best-fit values of Log$\epsilonstarUV$ corresponding roughly to the 2.3rd percentile of the Log$(\Mup / \Msun)$ marginalized posterior.
    The \textit{dotted, dark-red} and \textit{solid, dark-blue lines} further show the values of $\epsilonstarUV$ and $\Mup$ from the model of \citet{Cruz_2024} (as in Figure~\ref{fig:PopIII_UVLF_ref}), and the maximum halo mass hosting Pop~III stars at $z \approx 5.6 - 6.6$ from the \texttt{dustyGadget} simulation suite \citep{DiCesare_2023, Venditti_2023} as a reference. Note that either large stochasticities ($\sigmaUV \gtrsim 1.5$) or large high-mass cut-offs ($\Mup \gtrsim 10^{10} ~\Msun$, still well below the most massive Pop~III halo in the \texttt{dustyGadget} simulations) are required to fit the data within $2\sigma$.}
    \label{fig:PopIII_UVLF_corner_ref}
\end{figure}

Figure~\ref{fig:PopIII_UVLF_ref} shows the Pop~III UVLF resulting from the model of \citet{Cruz_2024} at $z = 6.1$ (with $\Delta z = 1$) when assuming a stochasticity parameter $\sigmaUV = 0.7$ (see Section~\ref{sec:methodology})\footnote{We assume values for the astrophysical parameters as in their table~I, with no LW feedback nor streaming velocity, a high-mass cut-off at the atomic-cooling threshold, and a UV enhancement $\varepsilon_\mathrm{\star,UV} / \varepsilon_\star = 2$.}, compared to the measurement from \citet{Fujimoto_2025} at $z \simeq 5.6 - 6.6$; note that the constraints from \citet{Fujimoto_2025} have been updated to also include the AMORE6 Pop~III candidate at $z \approx 5.7$ found by \citet{Morishita_2025} within the JOF field, with $\MUV \approx -14.5$. 
It is clear that, while traditional models can reproduce the Pop~II UVLF, the usual assumption of Pop~III star formation confined in molecular-cooling mini-halos -- as in \citet{Cruz_2024} -- falls short of recovering the Pop~III galaxy candidates GLIMPSE-16043 and AMORE6 at $\MUV \lesssim -14.5$, as the UVLF decays exponentially at $\MUV \lesssim -12$. A straightforward way to accommodate this bright candidates is to allow Pop~III star formation in heavier, atomic-cooling halos, as we illustrate with a model where the high-mass cut-off is extended to $\Mup \sim 10^{10} ~\Msun$ (``\textsc{Heavy}'' model), which succeeds in fitting the claimed detection as well as the upper limits.

In order to better quantify the high-mass cut-off requirements imposed by the data, in Figure~\ref{fig:PopIII_UVLF_corner_ref} we show the results of fitting the Pop~III UVLF model described in Section~\ref{sec:methodology} against the observational constraints; for details on our fitting strategy, refer to Appendix~\ref{app:model_details}. By considering the marginalized posterior for the $\Mup$ parameter when assuming $\sigmaUV = 0.7$, it is evident that a high-mass cut-off $\Mup \gtrsim 10^{10} ~\Msun$ is necessary to fit the observed point within $2\sigma$; this remains true when the uncertain GLIMPSE candidate is removed from the analysis -- provided that constraints on the total UVLF at the faint end are also taken into account --, while even larger values would be needed to also fit the bright candidate JOF-21739 at $\MUV \approx -17.6$ (see Appendix~\ref{app:alternative_datasets} for a discussion of alternative dataset combinations considered in this work). Therefore the data, if confirmed, may be hinting towards surviving Pop~III star formation in larger halos than commonly assumed.

This picture is supported by theoretical models. Simulations of Pop~III star formation in mini-halos (see e.g. \citealt{Bromm_2013} and references therein) commonly yield Pop~III masses of $\sim 10^2 - 10^3 ~\Msun$, which for a typical star-formation timescale of $\sim 10$~Myr correspond to SFRs of the order of $\sim 10^{-5} - 10^{-4} ~\si{\Msun.yr^{-1}}$. Calculations from \citet{Greif_Bromm_2006} and \citet{Greif_2008} further suggest that Pop~III star formation in atomic-cooling halos of $\Mh \sim 10^8 ~\Msun$ may yield up to $\sim 10^4 - 10^5 ~\Msun$ in stellar mass, resulting in even higher SFRs of $\sim 10^{-3} - 10^{-2} ~\si{\Msun.yr^{-1}}$. For a mini-halo and atomic-cooling mass scale of respectively $\Mh \sim 10^6 ~\Msun$ and $\Mh \sim 10^8 ~\Msun$, this would imply an upper limit on $\SFR / \Mh$ of $ \sim 10^{-11} - 10^{-10} ~\si{yr^{-1}}$, well above our \textsc{Heavy} model predictions; note that these limits correspond to an extreme scenario in which all halos are capable to form Pop~III stars, while in reality only a fraction of those halos will form Pop~IIIs (see e.g. the discussion in Appendix~\ref{app:model_details}). The model is also broadly in line with predictions from the \texttt{dustyGadget} \citep{Graziani_2020} cosmological simulation suite \citep{DiCesare_2023}: these are the largest simulations available (i.e. eight volumes of $\sim 70$~cMpc per side) that include a model for Pop~III star formation and feedback, thus representing a prime tool to explore Pop~III star formation in rare, massive halos\footnote{See \citealt{Venditti_2023, Venditti_2024_HeII, Venditti_2024_PISNe} for a discussion on the statistics of Pop~III star-forming environments during the EoR from the simulations and their detectability.}; however, refer to Appendix~\ref{app:model_details} for some cautionary notes on this comparison (particularly regarding contamination from co-existing Pop~II sources in the same halos), as well as more details on the $\SFR / \Mh$ estimate from the simulations.

Note that in the \texttt{dustyGadget} simulations, Pop~III stars can be found in host halos up to $\Mh \sim 10^{12} ~\Msun$ at these redshifts. In fact, even when the average SFR of the halo is dominated by Pop~II (which is usually the case in such large halos, see \citealt{Venditti_2023}), Pop~III stars can be found either in sub-regions at the periphery of the main galaxy or in small separate clumps/satellites. For example, a $\sim 10^{10} ~\Msun$ halo, forming a Pop\ III cluster of $\sim 10^5 ~\Msun$ over $\sim 5$~Myr (consistent with the candidate identified by \citealt{Fujimoto_2025}, and with the $\mathrm{SFR} \sim 0.3 ~\si{\Msun.yr^{-1}}$ inferred for AMORE6 from H$\beta$ measurements, see table~S1 of \citealt{Morishita_2025}) would yield $\SFR / \Mh \sim 2 \times 10^{-12} ~\si{yr^{-1}}$. With a large SFR-to-UV conversion enhancement with respect to Pop~II populations by a factor six (as postulated e.g. by \citealt{Liu_2025} for chemically homogeneous Pop~III stellar evolution models), this implies that $\sim 5\%$ of large halos hosting similar episodes of Pop\ III star formation would be sufficient to account for the observed Pop\ III UVLF.

\subsection{Role of stochasticity/burstiness}
\label{sec:z6.1_fit_bursty}

In the previous subsection, we showed that matching the bright Pop~III UV magnitudes reported by \citet{Fujimoto_2025} while retaining the Pop~II-inferred stochasticity value of $\sigmaUV$ requires Pop~III star formation to occur well into the atomic-cooling halo mass regime (as in the ``\textsc{Heavy}'' model of Figure~\ref{fig:PopIII_UVLF_ref}). However, brighter UV magnitudes can also be observed in low-mass halos that recently experienced a strong burst in star formation \citep{Mason_2023, Mirocha_Furlanetto_2023, Shen_2023, Sun_2023, Munoz_2023, Nikolic_2024, Gelli_2024, Chakraborty_Choudhury_2025}. In effect, a more stochastic Pop~III star-formation activity -- encoded by larger values of the $\sigmaUV$ parameter -- can alleviate the requirements on $\Mup$ by allowing an up-scattering of the Pop~III UV luminosity in low-mass halos.

In Figures~\ref{fig:PopIII_UVLF_ref} and~\ref{fig:PopIII_UVLF_corner_ref}, we demonstrate that a ``\textsc{Bursty}'' Pop~III star-formation model (assuming $\sigmaUV = 1.5$) can fit the data even with Pop~III stars confined within molecular-cooling halos; refer to Appendix~\ref{app:further_modeling} for a more extended discussion of the values of $\sigmaUV$ explored for this study. Note that, while the current data do not allow precise constraints on $\Mup$ when all parameters are allowed to vary, models with low values of both $\Mup$ and $\sigmaUV$ are strongly disfavored; larger values of $\sigmaUV \sim 1.2 - 1.8$ are also generally preferred (see e.g. Figure~\ref{fig:PopIII_UVLF_corner_fullposterior-subset} and its discussion in Appendix~\ref{app:further_modeling}).

Although the ``\textsc{Bursty}'' picture requires significantly larger stochasticity parameters than the constraints inferred from the UVLF at $z \approx 5.6 - 6.6$ ($\sigma_\mathrm{UV} \approx 0.7$, \citealt{Munoz_2023, Shuntov_2025}), such high values may not be unreasonable for this subdominant stellar population. In fact, Pop~III formation -- especially at late times -- is likely a more sporadic phenomenon (see e.g. \citealt{Venditti_2023}). Moreover, low-metallicity, Pop~III-forming clumps could more closely resemble higher-$z$ galaxies, which have been postulated to host highly stochastic star formation to account for the unexpected abundance of UV-bright sources at $z \gtrsim 10$. Particularly, values up to $\sigma_\mathrm{UV} \sim 3$ are in agreement with the observed UVLF at $z \gtrsim 10$ (\citealt{Finkelstein_2023, Perez-Gonzalez_2023, Harikane_2023, Donnan_2024}, see e.g. figures~2 and~3 of \citealt{Munoz_2023}), which would be largely consistent with our ``\textsc{Bursty}'' model. A mass-dependent stochasticity, with burstier low-mass halos due to their shallow potential well (in line with the prediction of hydrodynamical simulations, e.g., \citealt{Sun_2023}), can explain such an increase in $\sigmaUV$ up to $z \sim 12$ \citep{Gelli_2024}: in this model, a $\sigmaUV \gtrsim 1.5$ is assumed at $\Mh \sim 10^8 ~\Msun$, i.e., close to the atomic-cooling threshold at our redshifts of interest.

An important point to make is that our formalism considers the total UV magnitude arising from all Pop~III systems within each halo, without any sub-halo classification. If Pop~III stars are found in small satellite galaxies, these would naturally host burstier star-formation due to their sensitivity to feedback and environmental influences (see e.g. \citealt{Furlanetto_Mirocha_2022, Asada_2024} for a discussion of burstiness in interacting and non-interacting low-mass galaxies during the EoR). Therefore, our two proposed ``\textsc{Bursty}'' and ``\textsc{Heavy}'' models are not mutually exclusive, as Pop~III stars can be found not only in isolated mini-halos, but also in small, bursty sub-halos within larger, enriched halos. While our simple model for the Pop~III UVLF does not allow to discriminate between these scenarios, a study of the clustering properties of these systems, or of their nucleosynthetic signature \citep[e.g.,][]{Jeon_2014}, may shed further light on the conditions favoring Pop~III star formation at late times.

\section{A Pop~III sea in the ultra-deep Universe?}
\label{sec:deepMUV_high-z}

Both our ``\textsc{Heavy}'' and ``\textsc{Bursty}'' models predict a steep Pop~III UVLF faint-wards of the regime probed by current JWST surveys. For both these scenarios, the observation of GLIMPSE-16043 and AMORE6 implies a wealth of additional faint Pop~III systems to be discovered ($\gtrsim 10^{-2} ~\si{Mpc^{-3}.mag^{-1}}$ at $\MUV \gtrsim -14$). Our chances might also improve significantly by probing increasingly high redshifts, as Pop~III objects should become more prevalent with respect to their metal-enriched counterparts.

\begin{figure}
    \centering
    \includegraphics[width=\linewidth]{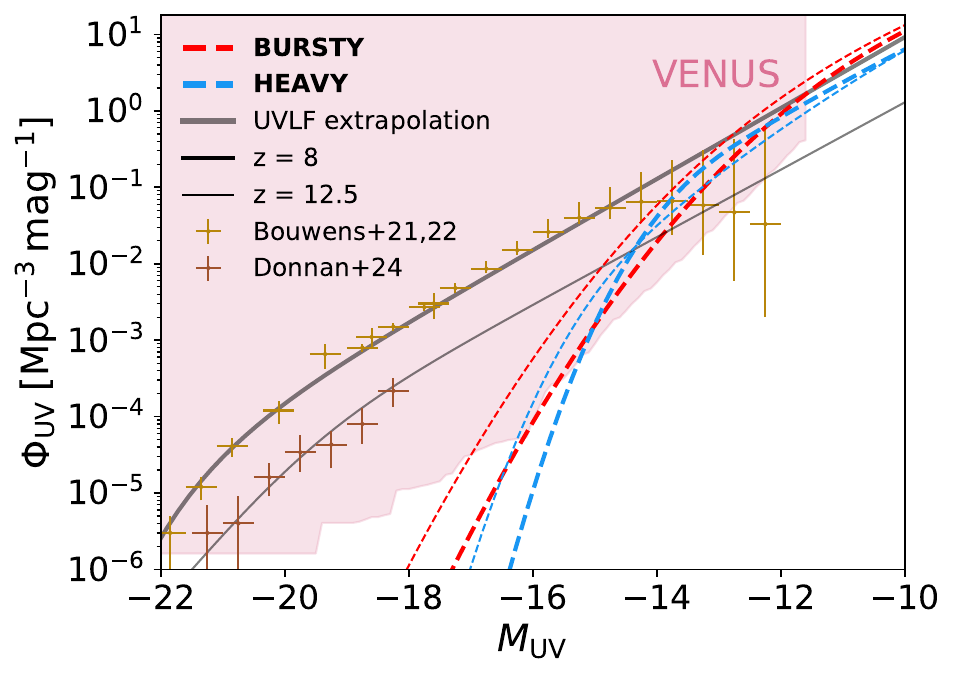}
    \caption{UVLFs resulting from the ``\textsc{Bursty}'' and ``\textsc{Heavy}'' models from Fig.~\ref{fig:PopIII_UVLF_ref} (\textit{red/blue, dashed lines}), if extrapolated to $z = 8$ and $z = 12.5$ (with progressively \textit{thinner linestyle}). 
    For comparison, we show the observed binned UVLF and fitting functions from \citet[][at $z = 8$]{Bouwens_2021} and \citet[][at $z = 12.5$]{Donnan_2024}, extrapolated towards fainter magnitudes. The bright end of the Pop~III UVLF shows a modest increase up to $z \approx 12.5$, while a strong decay of the total (Pop~II-dominated) UVLF is evident from observations. Lensing JWST surveys such as VENUS  \citep{Fujimoto_2025_VENUS} will allow deeper $M_\mathrm{UV}$ coverage (within the \textit{pink, shaded area}, computed for $z = 10$ with $\Delta z = 1$).}
    \label{fig:PopIII_UVLF_highz}
\end{figure}

To quantify these statements, we use our models to compute the predicted Pop~III UVLF at $z = 8$ and $z = 12.5$, shown in Figure~\ref{fig:PopIII_UVLF_highz}. We assume that the Pop~III SFE and the efficiency of SFR-to-UV luminosity conversion at given halo mass remain constant throughout this redshift range, so that the redshift evolution of the Pop~III UVLF is entirely determined by the underlying evolution of the halo mass function and of the accretion rate as a function of halo mass. This is obviously a simplification, as the exact physical mechanisms triggering late Pop~III star formation in large halos are still unclear, and possibly dependent on the assembly phase of the Universe, causing the SFE vs. $\Mh$ relation to also evolve over time. In \citet{Venditti_2023} we postulated that merger/interaction-driven star formation in pristine sub-clumps/satellites, or accretion of pristine gas from the large scale, may favor Pop~III star-formation in large halos, located in over-dense regions of the cosmic web (see e.g. Appendix~\ref{app:model_details}). However, further investigations are needed.

The Pop~III UVLF thus computed shows moderate increase with redshift at the bright end, while the faint end appears to be rather un-evolving, as the increase in the average SFR due to the more sustained accretion at high redshift is roughly balanced by the corresponding decrease in the total number of halos. Conversely, the total UVLF decays with redshift (see e.g. the constraints of \citealt{Bouwens_2021} at $z \approx 8$, with respect to those at $z \approx 12.5$ from \citealt{Donnan_2024}), hinting toward a larger contribution from Pop~III systems in the early Universe.
Note that the optimistic predictions of the ``\textsc{Bursty}'' model remain robust even after accounting for the effects of LW radiation and baryon–dark matter streaming velocities, which raise the threshold for Pop~III star formation in mini-halos (see Appendix~\ref{app:further_modeling} for further discussion). This resilience may stem from the larger relative number of low-mass halos at early times: in a similar context, a higher threshold is less impactful than the enhanced burstiness, which affects all halos uniformly. Moreover, the assumed value of $\sigmaUV = 1.5$ in the ``\textsc{Bursty}'' model is consistent with the overall star-formation activity observed at $z = 12.5$.

Predicting the Pop~III/II fraction as a function of redshift is extremely challenging, especially in the $M_\mathrm{UV}$ regime in which Pop~III actually starts dominating \citep[but see][]{Jaacks_2019}. Note that the fitting functions in Figure~\ref{fig:PopIII_UVLF_highz} are extrapolated well above the observed range, while the faint-end slope of the UVLF remains fairly unconstrained. Given the uncertainties involved in both the Pop~II and Pop~III faint-end UVLF predictions, we do not attempt to forecast the exact ratio of Pop~III vs. Pop~II systems over time. Nonetheless, we note that a top-heavy IMF component has been suggested in various studies to explain the abundant UV-bright galaxies at $z \gtrsim 10$ \citep[e.g.,][]{Inayoshi_2022, Finkelstein_2023, Harikane_2023, Harikane_2024, Yung_2024, Trinca_2024, Ventura_2024, Cueto_2024, Hutter_2025, Lu_2025, Harvey_2025, Jeong_2025, Mauerhofer_2025}; the study of \citet{Menon_2024} even suggests that both top-heavy IMFs and high star-formation efficiencies may be favored in conditions of low metallicity and high surface densities. Our predicted trends for the Pop~III vs. total UVLF in Figure~\ref{fig:PopIII_UVLF_highz} support a similar picture. This potentially indicates a significant Pop~III component in faint $z \gtrsim 10$ galaxies, and emphasizes the power of extending our Pop~III search to fainter magnitudes and higher redshifts. 

The JWST Cycle~4 treasury program, Vast Exploration for Nascent, Unexplored Sources (VENUS; \citealt{Fujimoto_2025_VENUS}) will probe a wide survey area around 60 lensing clusters, allowing us to constrain the UVLF down to a faint UV magnitude of $M_\mathrm{UV} \approx -12$ at $z \sim 10$, and potentially uncover many of these hidden Pop~III sources. Based on the effective survey volume of VENUS after the lens correction\footnote{
We calculate the effective survey volume as a function of $\MUV$ at $z = 10$ with $\Delta z = 1$, assuming the change is negligible in the specified redshift range from $z \approx 8$ to $z \approx 12.5$.
}, 
$\sim 14$ and~40 Pop~III systems are expected from our ``\textsc{Bursty}'' model respectively at $z = 8$ and $z = 12.5$, while $\sim 20$ Pop~III systems would be captured at $z = 8$ in the ``\textsc{Heavy}'' model, and $\sim 17$ at $z = 12.5$. We emphasize that higher values of $\sigmaUV$ are more likely at $z \gtrsim 10$, as discussed in Section~\ref{sec:z6.1_fit_bursty}, so that burstier models may be favored; also note that including the effect of the LW and streaming velocities (as in Appendix~\ref{app:further_modeling}), would only reduce our predicted number counts for the ``\textsc{Bursty}'' scenario by a factor $\sim 2$ and $\sim 1.5$ at $z = 8$ and $z = 12.5$ respectively (while the decrease for the ``\textsc{Heavy}'' model amounts to less than 5\%). Finally, by surveying an effective volume roughly three times larger than the programs included by \citet{Fujimoto_2025} -- over the same redshift range --, VENUS will reduce Poisson‐limited uncertainties on Pop~III number counts by a factor $\sim \sqrt{3}$, substantially tightening our constraints on the number density of these systems.

As a final remark, we note that the adopted strategy of \citet{Fujimoto_2025} of searching for Pop~III-dominated galaxy candidates is probably conservative. In fact, Pop~III stars may survive even in/around Pop~II-dominated galaxies \citep{Venditti_2023}, so that performing a similar search in a spatially resolved way may hold the potential for increased discoveries, enhancing our ability to discriminate between pure Pop~III systems and potential Pop~II contaminants.

\section{Summary and Conclusions}
\label{sec:conclusions}

Recent JWST observations at $z \approx 5.6 - 6.6$ have uncovered tentative metal-free (Pop~III) galaxy candidates with an unexpectedly bright UV magnitude, challenging conventional assumptions about the environments and efficiency of primordial star formation. These include AMORE6 within the JOF field at $z \approx 5.7$ (with $\MUV \approx -14.5$), GLIMPSE-16043 within the GLIMPSE field at $z \approx 6.5$ (with $\MUV \approx -15.9$), and an even brighter object within the JOF field at $z \approx 6.2$ (JOF-21739, with $\MUV \approx -17.6$). 
To interpret these surprising findings, we explored two contrasting yet complementary models, which are both able to reproduce the observed Pop~III UVLF by overturning one of our usual assumptions about Pop~III star formation:
(i) a ``\textsc{Heavy}'' scenario in which Pop~III stars continue forming efficiently above the atomic-cooling threshold; 
(ii) a scenario characterized by intense Pop~III star-formation activity in ``\textsc{Bursty}'' lower-mass halos.
Our analysis suggests that:
\begin{itemize}
    \item Pop~III star formation may not be limited to pristine, molecular-cooling mini-halos (with a typical mass $\Mh \lesssim 10^8 ~\Msun$ at $z \simeq 5.6 - 6.6$). Halos up to a high-mass cut-off $\Mup \gtrsim 10^{10} ~\Msun$ (or even $\Mup \gtrsim 10^{12} ~\Msun$, if the constraint from JOF-21739 is taken into account) are in fact viable sites for late Pop~III formation.
    \item Strong burstiness in the Pop~III star-formation activity can likewise explain the data, provided the scatter in UV luminosity output of these halos is large enough. Particularly, values of $\sigmaUV \gtrsim 1.5$ are required to be consistent with AMORE6 and GLIMPSE-16043 with Pop~III star formation confined in mini-halos (or even $\sigmaUV \gtrsim 2$ when including JOF-21739), much larger than the typical values inferred for the average star formation at these redshifts ($\sigma_\mathrm{UV} \sim 0.7$) -- albeit consistent with the values derived at $z \gtrsim 10$.
\end{itemize}

Both scenarios hint towards a larger, hidden Pop~III population: if current candidates lie near the bright end of the Pop~III UVLF, a steep faint-end slope suggests many more systems may exist just below current detection thresholds. The evolution of the Pop~III UVLF in our models indicates that by pushing the redshift frontier even further, the abundance of Pop~III systems and their relative contribution to the total UVLF may both increase, significantly strengthening our detection capabilities. Moreover, such an enhancement of the density of Pop III systems is likely to impact the timing of reionization and cosmic dawn~\citep{Munoz_2022}, as well as the signal from compact Pop~III remnants (e.g. gravitational waves and the X-ray background).
These observables, together with galaxy clustering, additionally provide promising tools to disentangle whether Pop III star formation is ``\textsc{Bursty}'' or ``\textsc{Heavy}'', and in the former case it can be enhanced by a halo-occupation distribution model~\citep{Berlind_Weinberg_2002}.

While we cannot draw definitive conclusions about the sites of high-redshift Pop~III formation from the present analysis alone, the formalism developed here offers a powerful and versatile framework to interpret upcoming deep-field surveys. In particular, it will allow future works to trace the redshift evolution of the Pop~III star-formation efficiency and its connection to halo mass, ultimately offering new insights into the final chapters of primordial star formation.\\

%% IMPORTANT! The old "\acknowledgment" command has be depreciated. It was
%% not robust enough to handle our new dual anonymous review requirements and
%% thus been replaced with the acknowledgment environment. If you try to 
%% compile with \acknowledgment you will get an error print to the screen
%% and in the compiled pdf.
%% 
%% Also note that the akcnowlodgment environment does not support long amounts of text. If you have a lot of people and institutions to acknowledge, do not use this command. Instead, create a new \section{Acknowledgments}.

%\begin{acknowledgments}

AV acknowledges funding from the Cosmic Frontier Center and the University of Texas at Austin’s College of Natural Sciences.
JBM was supported by NSF Grants AST-2307354 and AST-2408637, and by the NSF-Simons AI Institute for Cosmic Origins.
This research was also supported in part by grant NSF PHY-2309135 to the Kavli Institute for Theoretical Physics (KITP).

%\end{acknowledgments}

%% To help institutions obtain information on the effectiveness of their 
%% telescopes the AAS Journals has created a group of keywords for telescope 
%% facilities.
%
%% Following the acknowledgments section, use the following syntax and the
%% \facility{} or \facilities{} macros to list the keywords of facilities used 
%% in the research for the paper.  Each keyword is check against the master 
%% list during copy editing.  Individual instruments can be provided in 
%% parentheses, after the keyword, but they are not verified.

%% Similar to \facility{}, there is the optional \software command to allow 
%% authors a place to specify which programs were used during the creation of 
%% the manuscript. Authors should list each code and include either a
%% citation or url to the code inside ()s when available.

\software{
\texttt{\href{https://github.com/JulianBMunoz/Zeus21}{Zeus21}} \citep{Munoz_2023_Zeus21},
\texttt{\href{https://cobaya.readthedocs.io/en/latest/theory_class.html}{classy}} \citep{Blas_2011_classy},
\texttt{dustyGadget} \citep{Graziani_2020},
\texttt{\href{https://dan.iel.fm/emcee}{emcee}} \citep{Foreman-Mackey_2013_emcee},
\texttt{\href{https://github.com/dfm/corner.py}{corner}}, \citep{Foreman-Mackey_2016_corner},
\texttt{\href{https://numpy.org}{numpy}} \citep{VanDerWalt_2011_numpy, Harris_2020_numpy},
\texttt{\href{https://matplotlib.org}{matplotlib}} \citep{Hunter_2007_matplotlib}, 
\texttt{\href{https://scipy.org}{scipy}} \citep[\href{https://mail.python.org/pipermail/python-list/2001-August/106419.html}{Jones et al. 2001};][]{Virtanen_2020_scipy}.
A simplified version of the notebook and the modified \texttt{Zeus21} code used for the analysis can be found in the public Zenodo repository of \citet{Venditti_2025_notebook}.
}

%% Appendix material should be preceded with a single \appendix command.
%% There should be a \section command for each appendix. Mark appendix
%% subsections with the same markup you use in the main body of the paper.

%% Each Appendix (indicated with \section) will be lettered A, B, C, etc.
%% The equation counter will reset when it encounters the \appendix
%% command and will number appendix equations (A1), (A2), etc. The
%% Figure and Table counter will not reset.

%% For this sample we use BibTeX plus aasjournals.bst to generate the
%% the bibliography. The sample631.bib file was populated from ADS. To
%% get the citations to show in the compiled file do the following:
%%
%% pdflatex sample631.tex
%% bibtext sample631
%% pdflatex sample631.tex
%% pdflatex sample631.tex

\appendix

\section{Pop~III SFE/UVLF model and simulations}
\label{app:model_details}

The Pop~III SFE and UVLF models described in Section~\ref{sec:methodology} are computed by adopting a modified version of the publicly available code \texttt{Zeus21}\footnote{\url{https://github.com/JulianBMunoz/Zeus21}} \citep{Munoz_2023_Zeus21}, accounting for our updates to the Pop~III model of \citet{Cruz_2024}. The model is fit via Markov Chain Monte Carlo sampling using the public Python package \texttt{emcee} \citep{Foreman-Mackey_2013_emcee}. We employ 50 walkers each taking 5000 steps, discard the first 1000 as burn‐in, and thin the remaining chain by a factor of ten to reduce autocorrelation. Flat priors are imposed on $\mathrm{Log}\epsilonstar \in [-5,-1]$ and $\mathrm{Log}(\Mup/\Msun) \in [8,15]$, and -- when varied (Appendix~\ref{app:further_modeling}) -- on $\betastar \in [-2.0,0.5]$ and $\sigmaUV \in [0.3,3]$. Number‐count uncertainties are derived from single‐sided Poisson $1\sigma$ limits for zero or one event \citep{Gehrels_1986}, and a Poisson likelihood is adopted for both detections and upper limits, although we note that departures from a Gaussian approximation are mostly confined within large $\Mup \gtrsim 10^{12} ~\Msun$, well outside our parameter region of interest. We estimated the impact of cosmic variance using the analytic formalism of \citet{Trenti_Stiavelli_2008}, adopting the effective survey volume at the intrinsic UV luminosity appropriate for each of our candidates. The resulting fractional uncertainties are $\sim 27\%$, 25\% and 20\% in the $\MUV$ bins including AMORE6, GLIMPSE-16043 and JOF-21793 respectively, which are sub-dominant compared to the Poisson errors given the small-number statistics. These estimates are largely independent of the detailed Pop~III star-formation scenarios (e.g., bursty histories), since cosmic variance primarily depends on survey geometry and the clustering bias of host halos. We caution that models invoking more massive host halos would correspond to higher halo bias: for example, we find a factor $\sim 1.6$ between our \textsc{Heavy} and \textsc{Bursty} models, which would translate into a corresponding increase of the cosmic variance uncertainty; however, we verified that the bias associated with Pop~III hosts in our reference models is always lower than the bias resulting from all galaxy populations (as in our fiducial estimate), as even in models with relatively high $\Mup$ (e.g. our \textsc{Heavy} model, with $\Mup \approx 10^{10} ~ \Msun$), Pop~III stars still predominantly reside in lower-mass hosts than the average stellar populations at this redshift.

To compare our results with predictions from the \texttt{dustyGadget} simulation suite (e.g. bottom-left panel of Figure~\ref{fig:PopIII_UVLF_ref}), we computed the fraction of Pop~III halos in bins of halo mass within six of the \texttt{dustyGadget} cubes, and the corresponding average Pop~III SFR per unit halo mass among all Pop~III-hosting halos in each bin; the product of these two quantities is shown in the plot as a function of the average $\Mh$ in each bin. The Pop~III SFR in a given halo is estimated as the total mass of currently alive Pop~III stars divided by the average Pop~III lifetime with the assumed $[100, 500] ~\Msun$ Salpeter IMF ($\sim 3$~Myr), and the mass is rescaled by a factor $\etaIII = 0.3$ with respect to the Pop~III mass resolution element $M_\mathrm{III,res} \sim 2 \times 10^6 ~\Msun$, as in \citet{Venditti_2024_HeII, Venditti_2024_PISNe}. The binning size (demonstrated by the $x$-axis error-bars) are chosen in order to have an approximately even number of halos per bin. The uncertainty on $\SFR/\Mh$ among Pop~III halos is inferred from its standard deviation within each bin, and we further assume a Poisson variance on the number of Pop~III halos (with their propagation resulting in the $y$-axis error-bars).

While the SFR per unit halo mass among Pop~III halos decreases steadily with $\Mh$, the increasing Pop~III fraction above $\Mh \sim 10^{10} ~\Msun$ causes a flattening in the relation between the Pop~III SFE in all halos. In such larger halos, located in over-dense regions of the cosmic web, specific physical conditions such as a larger infall of pristine gas from the large scale and/or a higher level of galaxy interactions (causing compression in the gas) may trigger late Pop~III star formation in the outskirts of the central galaxies or in their satellites; see e.g. the discussion of \citealt{Venditti_2023}. Note that the {uncertainty on the $\etaIII$ parameter (i.e. on the total Pop~III mass formed in a single star-formation episode) also absorbs the underlying uncertainty on the UV boost $\epsilonstarUV/\epsilonstar \equiv \kappaUV / \kappa_\mathrm{UV,ref}$ for top-heavy IMFs, so that the somewhat extreme choice of $\etaIII = 0.3$ in the plot potentially incorporates the effect of larger UV luminosities per unit mass. Regardless, Pop~III halos in the simulations are found up to large halo masses of $\sim 10^{12} ~\Msun$ (see the dark-blue, solid lines e.g. in the bottom-left panel of Figure~\ref{fig:PopIII_UVLF_corner_ref}), i.e. well into the atomic-cooling regime (above the red, dotted lines).

As emphasized in Section~\ref{sec:z6.1_fit_bursty}, our formalism considers the total UV magnitude arising from Pop~III systems within a halo of mass $\Mh$, without taking into account any sub-halo classifications. A similar strategy has been adopted when computing the SFE from the \texttt{dustyGadget} simulations, to ensure a homogeneous comparison. Although every Pop~III–hosting halo in the mass range considered also contains a dominant Pop~II component -- which would certainly yield clear observational signatures such as strong metal lines --, pristine gas pockets capable of forming Pop~III stars should remain spatially segregated from Pop II star-forming regions in order to avoid metal pollution. Such isolated Pop III-forming pockets may, in principle, be identified as individual sources in surveys like that of \citet{Fujimoto_2025}. A detailed analysis of Pop~II contamination and of the role of segregated sub-structures in driving late Pop~III star formation in massive halos lies beyond the scope of this work. However, \citet{Venditti_2023} offer preliminary evidence for physical separation between the two populations: their figure~10, for example, compares the radial distribution of Pop~III stellar populations relative to the halo center of mass with the overall stellar mass–weighted radius.

\section{Alternative datasets}
\label{app:alternative_datasets}

\begin{figure*}
    \centering
    \includegraphics[width=0.49\linewidth]{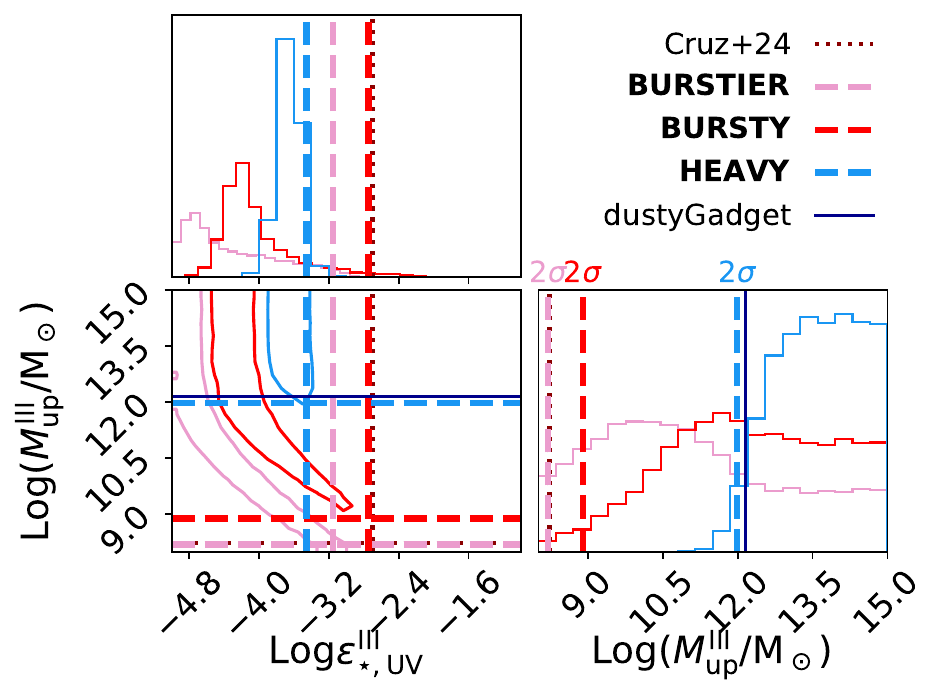}\\
    \includegraphics[width=0.49\linewidth]{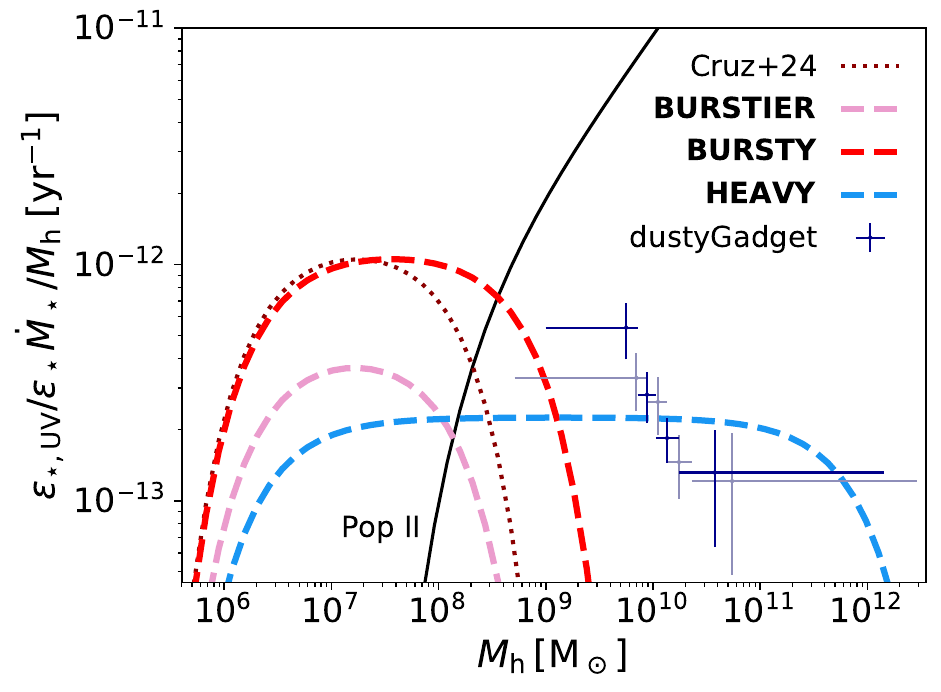}
    \includegraphics[width=0.49\linewidth]{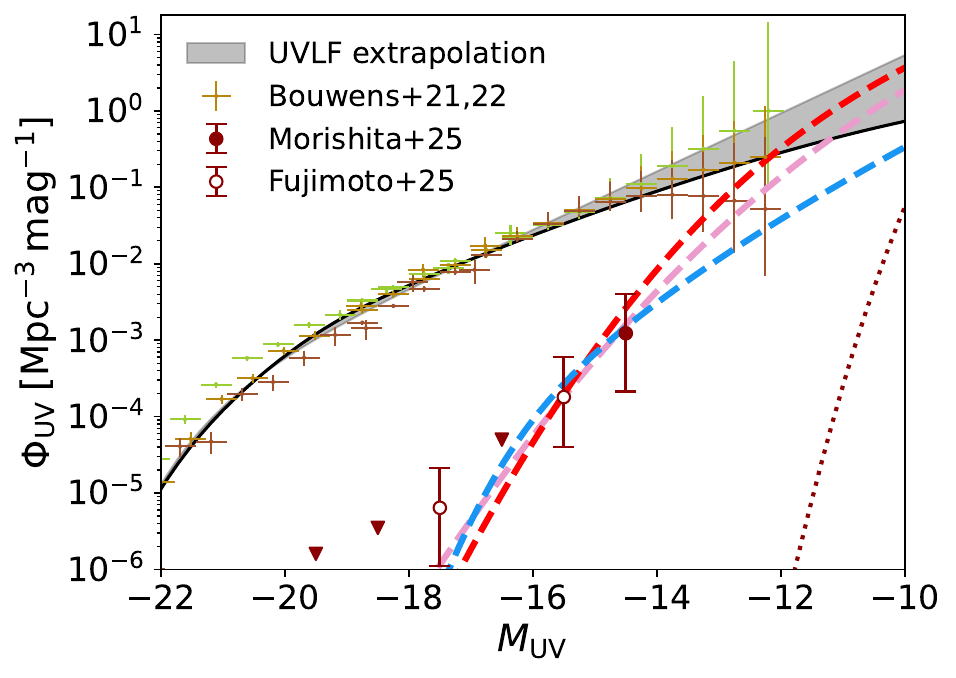}
    \caption{Same as Figures~\ref{fig:PopIII_UVLF_ref} and~\ref{fig:PopIII_UVLF_corner_ref}, but also including the uncertain JOF-21739 Pop~III candidate at bright $\MUV \approx -17.6$ in the fit. The ``\textsc{Bursty}''/``\textsc{Heavy}'' models in this case correspond to larger $\Mup \sim 10^9 / 10^{12} ~\Msun$, and $\epsilonstarUV \sim 10^{-2.8} / 10^{-3.4}$ (with the usual reference $\sigmaUV = 1.5/0.7$). An additional $\sigmaUV = 2$ scenario is included in \textit{pink}, which approximately corresponds to the minimum stochasticity allowing to fit the data within $2\sigma$ with a cut-off close to the atomic-cooling halo mass threshold; see e.g. the ``\textsc{Burstier}'' model ($\Mup \sim 10^8 ~\Msun$, $\epsilonstarUV \sim 10^{-3.1}$).}
    \label{fig:PopIII_UVLF_ref+JOF}
\end{figure*}

Figure~\ref{fig:PopIII_UVLF_ref+JOF} illustrates the results of fitting the model described in Section~\ref{sec:methodology} by also including the tentative candidate JOF-21739 at a bright $\MUV \sim -17.6$. We see that, while GLIMPSE-16043 and AMORE6 only require $\Mup \gtrsim 10^{10} ~\Msun$ (when assuming $\sigmaUV = 0.7$), a much higher cut-off $\Mup \gtrsim 10^{12} ~\Msun$ is needed to fit JOF-21739. This pushes the model beyond the limit of the maximum Pop~III halo mass found within the \texttt{dustyGadget} simulation suite. Moreover, even with a large $\sigmaUV = 1.5$, the increased burstiness cannot account for this bright candidate when Pop~III star formation is assumed to be confined within mini-halos ($\Mup \gtrsim 10^9 ~\Msun$ is necessary to fit the data within $2\sigma$). Even larger values ($\sigmaUV \gtrsim 2$) are needed in order to soften the $\Mup$ constraint enough (i.e. to $\Mup \gtrsim 10^8 ~\Msun$) to justify a similar picture.

\begin{figure*}
    \centering
    \includegraphics[width=0.49\linewidth]{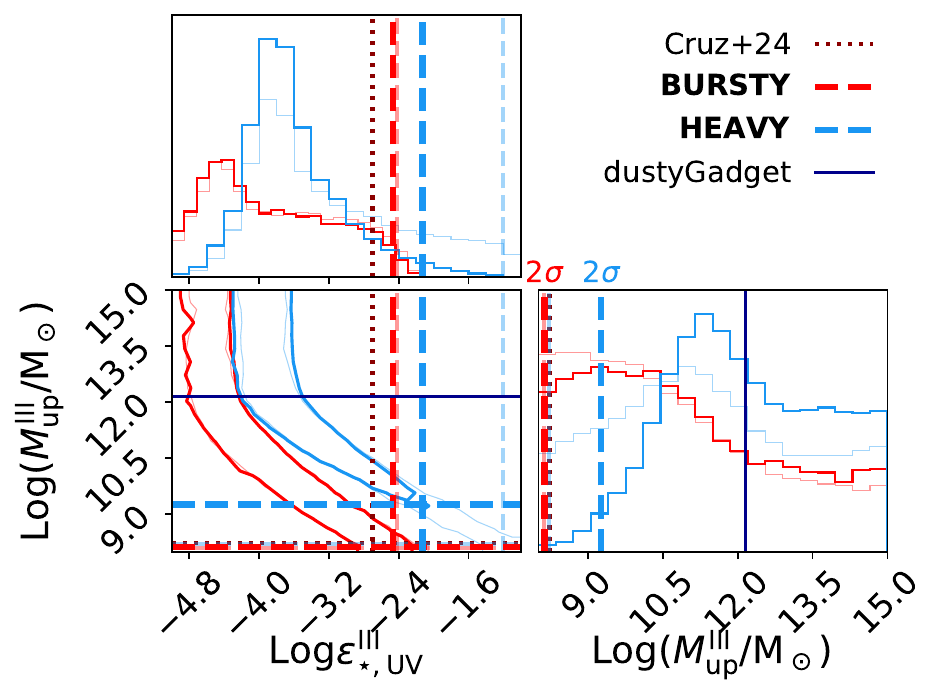}\\
    \includegraphics[width=0.49\linewidth]{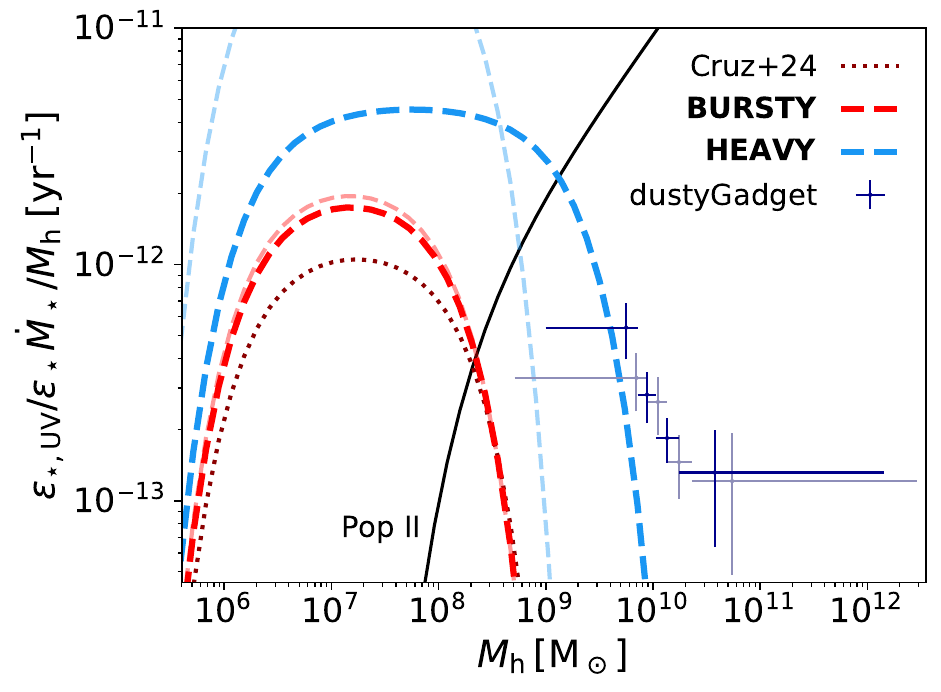}
    \includegraphics[width=0.49\linewidth]{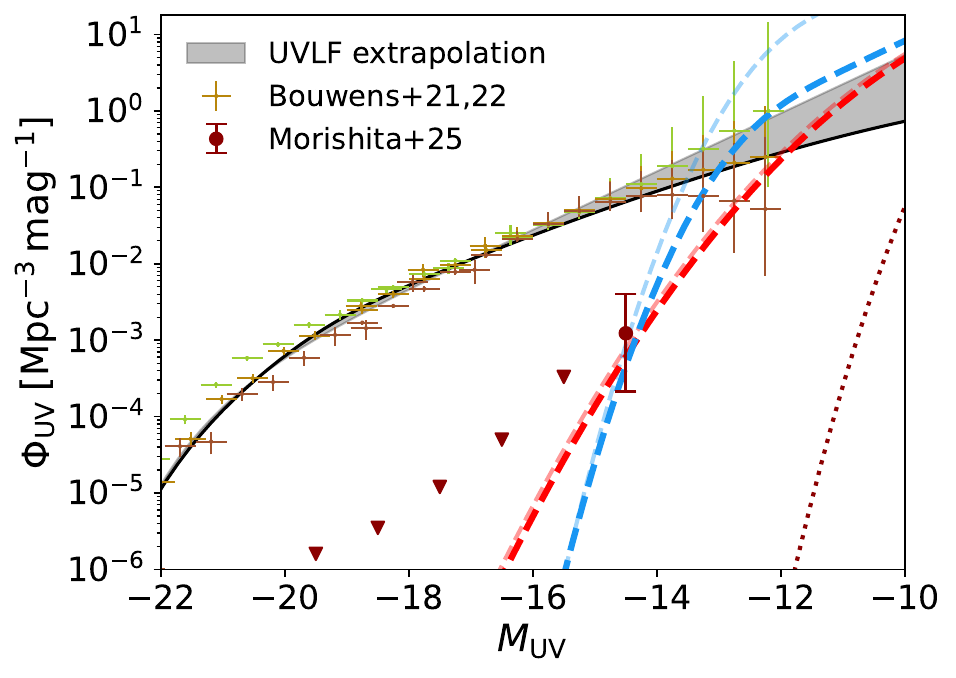}
    \caption{Same as Figures~\ref{fig:PopIII_UVLF_ref} and~\ref{fig:PopIII_UVLF_corner_ref}, but only including the AMORE6 Pop~III candidate from \citet{Morishita_2025} and the $z = 6$ total UVLF constraints from \citet{Bouwens_2022} in the fit (as detailed in the text); \textit{shaded lines} refer to the results of corresponding fits without the inclusion of the total UVLF constraints. While the latter allow $\Mup \sim 10^8 ~\Msun$ with $\epsilonstarUV \sim 10^{-2.4}$ (for the ``\textsc{Bursty}'' model) and a very large $\epsilonstarUV \sim 10^{-1.2}$ (for the ``\textsc{Heavy}'' model), values of $\Mup \sim 10^8 / 10^{9.5} ~\Msun$ and $\epsilonstarUV \sim 10^{-2.5} / 10^{-2.1}$, closer to the case including GLIMPSE-16043 (Figures~\ref{fig:PopIII_UVLF_ref} and~\ref{fig:PopIII_UVLF_corner_ref}) are found for the ``\textsc{Bursty}''/``\textsc{Heavy}'' models when the total UVLF constraints are included.}
    \label{fig:PopIII_UVLF_ref_AMORE6}
\end{figure*}

As evidence of [OIII] has been reported for GLIMPSE-16043, raising a caution on the adopted photometric selection criterion (Fujimoto et al. in prep., private communication), we repeated the analysis by solely relying on the more robust spectroscopic candidate AMORE6. The results are shown in Figure~\ref{fig:PopIII_UVLF_ref_AMORE6}. Constraints on the $z = 6$ total UVLF at $M_\mathrm{UV} \gtrsim -14.5$ from \citet{Bouwens_2022} have also been provisionally included in the fit, assuming Pop~II model parameters from \citet{Munoz_2023}. Note that, a fully self-consistent approach would require a simultaneous fit over the combined Pop III/II parameter space -- leveraging both the total UVLF constraints from \citet{Bouwens_2021, Bouwens_2022} and the Pop~III UVLF bounds from \citet{Fujimoto_2025, Morishita_2025} --, while our goal here is simply to illustrate the qualitative impact of the total UVLF constraints on our results, rather than derive precise parameter values.

Incorporating the total UVLF data pushes the preferred $\Mup$ to higher values than in the AMORE6-only fit, in order to avoid overproducing the faint end of the overall UVLF. This adjustment brings the $\Mup$ posterior into closer agreement with the case where both AMORE6 and GLIMPSE-16043 are included (Figure \ref{fig:PopIII_UVLF_corner_ref}); in fact, although $\Mup \sim 10^8,\Msun$ remains allowed when fitting only AMORE6, a similar value in the ``\textsc{Heavy}'' scenario would demand an implausibly high $\epsilonstarUV \sim 10^{-1.2}$. In other words, even though the selection criterion of \citet{Fujimoto_2025} has been put into question by recent observations, excluding the photometrically-selected candidates GLIMPSE-16043 and JOF-21793 but enforcing the total UVLF constraints at the faint end still yields similar conclusions: namely, all these bright candidates within the considered survey volumes are surprising in light of traditional Pop~III models, and either large host halo masses or bursty star formation histories (in line with our ``\textsc{Heavy}''/``\textsc{Bursty}'' models) should be invoked to explain the observations.
Finally, note that adding GLIMPSE-16043 has only a marginal impact when the bright JOF-21793 source is included as in Figure \ref{fig:PopIII_UVLF_ref+JOF}, as the latter strongly dominates the fit.

\section{Further modeling}
\label{app:further_modeling}

\begin{figure*}
    \centering
    \includegraphics[width=0.49\linewidth]{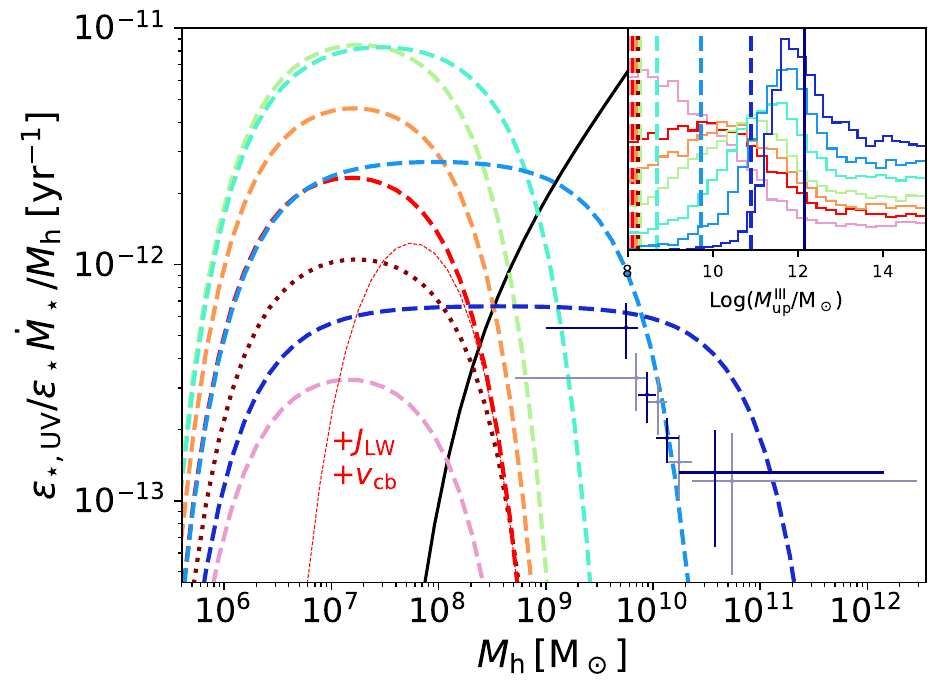}
    \includegraphics[width=0.49\linewidth]{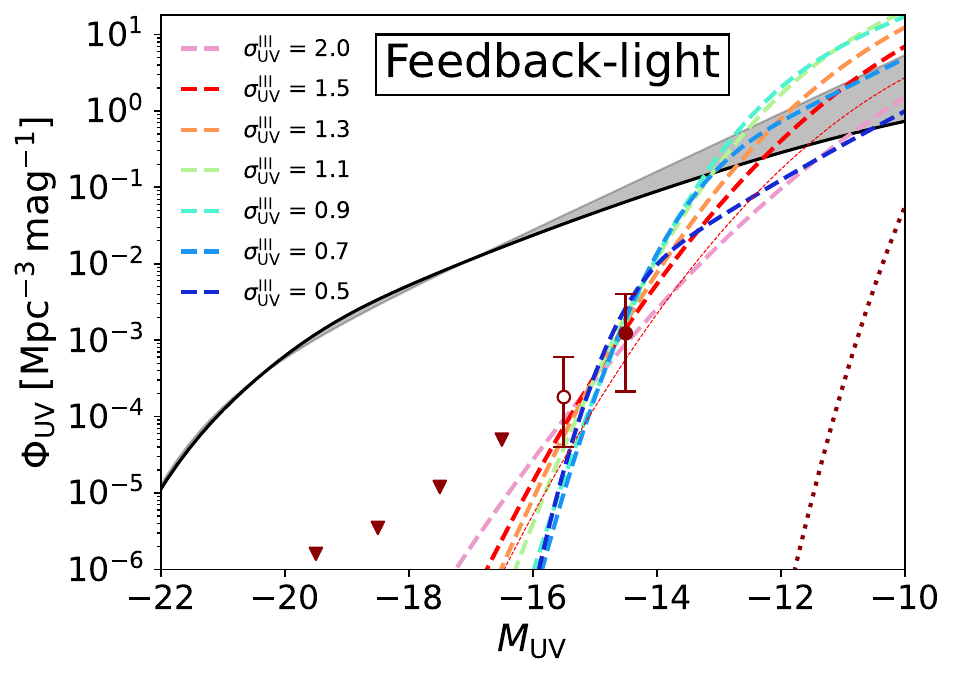}
    \includegraphics[width=\linewidth]{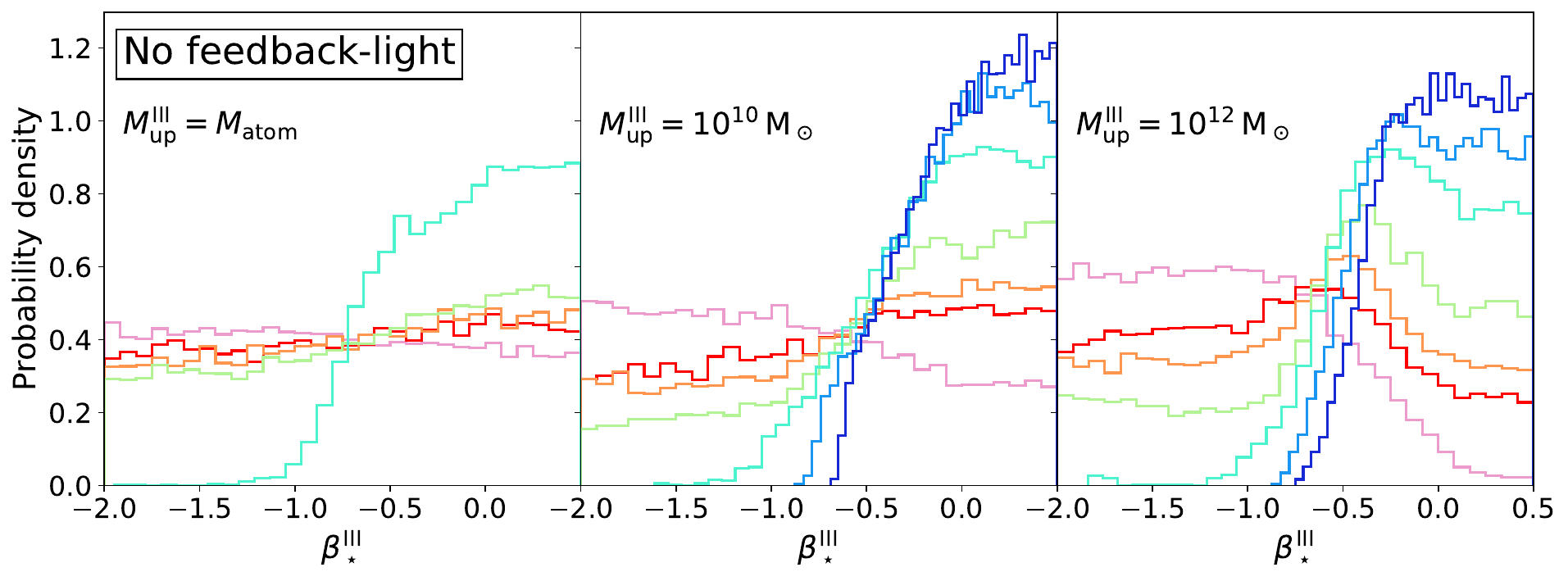}
    \caption{\textbf{Top:} same as Figure~\ref{fig:PopIII_UVLF_ref}, but exploring different values for the stochasticity parameter $\sigmaUV$ (from 0.5 to 2, in \textit{rainbow color-scale}); the \textbf{top-right inset} in the \textbf{left panel} shows the 1D posterior for the Log$(\Mup/\Msun)$ parameter corresponding to each chosen value of $\sigmaUV$, as in the bottom-left panel of Figure~\ref{fig:PopIII_UVLF_corner_ref}. More stringent requirements are imposed on $\Mup$ for lower values of $\sigmaUV$, ranging from $\Mup \gtrsim 10^{8} ~\Msun$ to $\Mup \gtrsim 10^{11} ~\Msun$ in order to fit the data within $2\sigma$. The \textit{thin, dashed, red line} shows a variation of the ``\textsc{Bursty}'' model including the effect of the streaming velocities and LW ($J_\mathrm{LW}$ and $v_\mathrm{cb}$, see text for more details), which produces a slight decrease in the overall UVLF. \textbf{Bottom:} 1D posterior for the $\betastar$ parameter resulting from fits with fixed $\Mup = M_\mathrm{atom} / 10^{10} ~\Msun / 10^{12} ~\Msun$ (\textbf{left/center/right}) and free $\epsilonstarUV$ and $\betastar$, for different values of $\sigmaUV$ (same \textit{color-scale} as the top panels), showing that our standard assumption of $\betastar = 0$ (``feedback-light'' model, as in the top panels) is consistent with these results, albeit slightly disfavored in models with both large $\Mup$ and large $\sigmaUV$.}
    \label{fig:PopIII_UVLF_further_modeling}
\end{figure*}

We examine the implications of different scenarios for the star-formation burstiness in our models, encoded by the stochasticity parameter $\sigmaUV$. The top panels of Figure~\ref{fig:PopIII_UVLF_further_modeling} show the results of fitting our model against the data with different $\sigmaUV$s with respect to the reference case $\sigmaUV = 0.7$, which would imply $\Mup \gtrsim 10^{10} ~\Msun$ (as in the ``\textsc{Heavy}'' model of Figure~\ref{fig:PopIII_UVLF_ref}). As we discussed in Section~\ref{sec:z6.1_fit_bursty}, larger values of $\sigmaUV$ can in fact soften the constraints on $\Mup$ by allowing an up-scattering of the Pop~III UV luminosity in low-mass halos. While assuming a tighter relation between $\LUV$ and $\Mh$ (e.g. $\sigmaUV = 0.5$) requires a higher cut-off in order to fit the data ($\Mup \gtrsim 10^{11} ~\Msun$), more scattered relations allow lower cut-offs (e.g. $\Mup \gtrsim 10^{8} - 10^{8.5} ~\Msun$ for $\sigmaUV$ between 0.9 and 1.3), albeit with a large efficiency ($\epsilonstarUV \sim 10^{-2}$). 
An even larger value of $\sigmaUV = 1.5$ finally pushes the model close to the atomic-cooling threshold, with a normalization $\epsilonstarUV$ consistent with a UV-boost by a factor $\epsilonstarUV / \epsilonstar \sim 5$ with respect to the model of \citet{Cruz_2024} (as in the ``\textsc{Bursty}'' model of Figure~\ref{fig:PopIII_UVLF_ref}). 

So far we have neglected both LW radiation and streaming velocities, as their impact on the molecular-cooling limit mass $\Mmol$ (in Equation~\ref{eq:fduty_III}) is mostly limited to the faint end of the UVLF. We tested this assumption by exploring the effect of streaming velocities and of a LW background consistent with the ``Pop~II'' + ``Pop~III, no feedback'' model of \citet{Cruz_2024} (see the left panel of their figure~3). A small decrease in the overall Pop~III UVLF is evident over the whole UV magnitude range for the ``\textsc{Bursty}'' model when these two quantities are included (as shown in the top panels of Figure~\ref{fig:PopIII_UVLF_further_modeling}); this is due to the lower contribution of both faint low-mass halos and low-mass halos up-scattered to the bright end. On the other hand, we find no appreciable difference in the ``\textsc{Heavy}'' model above $\MUV \approx -10$. This is true over the entire redshift range explored in the present work (i.e. up to $z \approx 12.5$), while the effect on the overall Pop~III abundances of the ``\textsc{Bursty}'' model remains moderate.

Another assumption of our model is that of a high-mass-end slope of the SFE $\betastar = 0$ (see Equation~\ref{eq:fstar_III}), which is expected if Pop~III stars formed too fast for feedback to become effective in quenching the star formation (``feedback-light'' model). In the bottom panel of ~\ref{fig:PopIII_UVLF_further_modeling} we show the results of various fits (for different values of $\sigmaUV$) in which $\betastar$ is allowed to vary -- together with $\epsilonstarUV$ --, with fixed $\Mup = \Matom, 10^{10} ~\Msun$ and~$10^{12} ~\Msun$. We see that for small $\sigmaUV$s, positive values of $\betastar$ are favored, while as $\sigmaUV$ increases, the distribution of $\betastar$ progressively flattens and becomes uninformative, with this effect more pronounced for the lowest $\Mup$; conversely, at the highest $\Mup$, increasingly large $\sigmaUV$ values lead to a growing preference against $\betastar > 0$, albeit $\betastar = 0$ remains marginally consistent with the data.

\begin{figure*}
    \centering
    \includegraphics[width=0.49\linewidth]{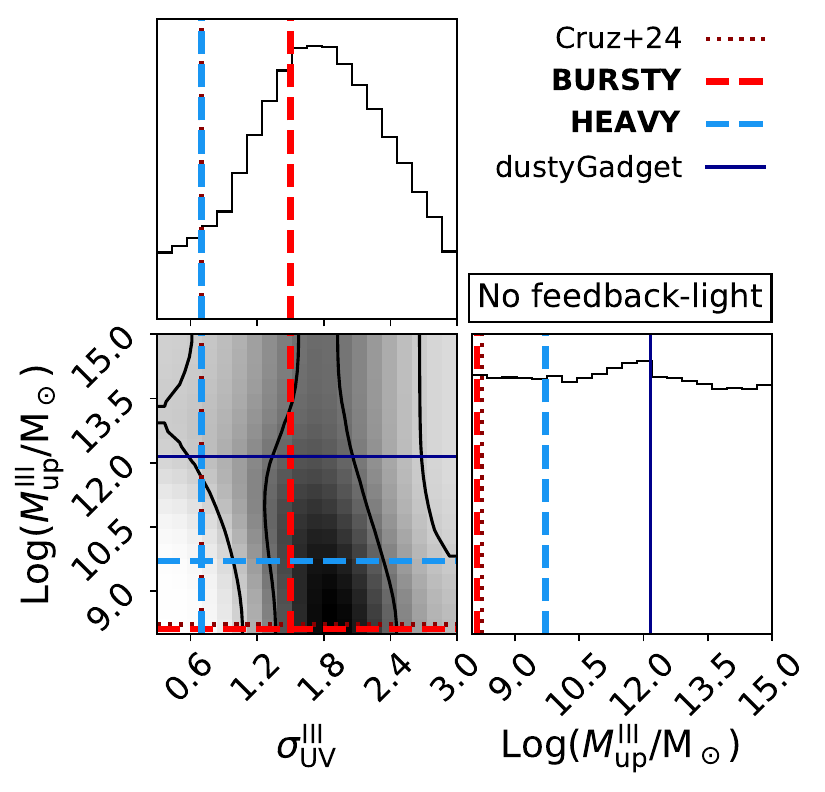}
    \includegraphics[width=0.49\linewidth]{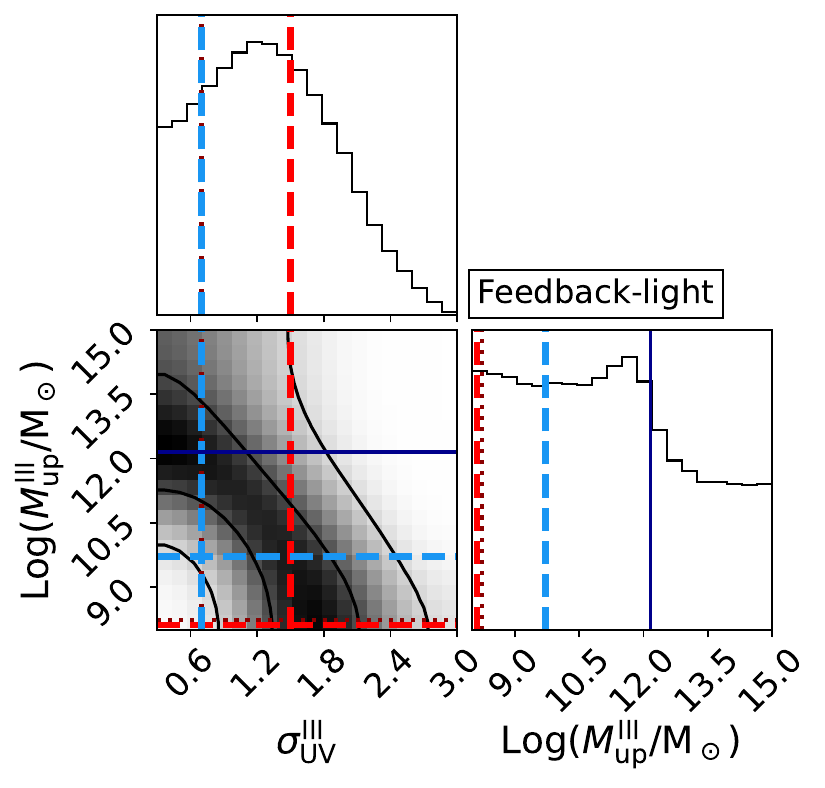}
    \caption{Joint posterior for the stochasticity parameter $\sigmaUV$ and the high-mass cut-off $\Mup$, resulting by fitting our Pop~III UVLF model against the observed data points from \citet{Fujimoto_2025} and \citet{Morishita_2025} at $z \approx 5.6 - 6.6$, when all model parameters ($\sigmaUV$, $\epsilonstarUV$, $\betastar$ and $\Mup$) are allowed to vary (\textbf{left}) and when we fix $\betastar = 0$ (\textbf{right}); 1$\sigma$ and 2$\sigma$ contour lines are shown as \textit{black, solid lines} in the 2D posterior. Reference values for our ``\textsc{Bursty}'' and ``\textsc{Heavy}'' models, for the model of \citet{Cruz_2024} and for the \texttt{dustyGadget} simulation suite are included as in Figure~\ref{fig:PopIII_UVLF_corner_ref}. Low values of both $\sigmaUV$ and $\Mup$ simultaneously are strongly disfavored, even more so when $\betastar$ is free to vary.}
    \label{fig:PopIII_UVLF_corner_fullposterior-subset}
\end{figure*}

In the left panel of Figure~\ref{fig:PopIII_UVLF_corner_fullposterior-subset} we present the $\Mup$ and $\sigmaUV$ posterior resulting from a fit exploring the full parameter space (i.e. with $\sigmaUV$, $\epsilonstarUV$, $\betastar$ and $\Mup$ all free to vary). The marginal distribution of $\Mup$ appears essentially flat, demonstrating the limited constraining power of current data on the precise value of $\Mup$. Nonetheless, examining the 2D posterior in the $\sigmaUV - \Mup$ plane reveals that the region of low $\Mup$ and low $\sigmaUV$ is strongly disfavored. Note that, while the ``\textsc{Bursty}'' model appears favored as it falls well within 1$\sigma$ confidence limits, the ``\textsc{Heavy}'' model lies outside the 2$\sigma$ boundary and would require even larger $\Mup$ ($\sim 10^{11} ~\Msun$) to fit the data, when $\betastar$ is allowed to vary; if instead one fixes $\betastar = 0$ as in our reference models (right panel of Figure~\ref{fig:PopIII_UVLF_corner_fullposterior-subset}), the ``\textsc{Heavy}'' curve skirts the 2$\sigma$ boundary.

These results are in line with our main conclusion, that to reproduce the observations with host masses close to the atomic-cooling limit one must invoke burstier star-formation histories (i.e. higher $\sigmaUV$s), whereas if the star-formation variability is mild larger host masses are required (higher $\Mup$). We remark that the goal of this paper is to present a theoretical framework exploring the potential implications of bright Pop III systems at late times, while precise parameter estimation will require tighter observational constraints.

\bibliography{main}{}

\begin{thebibliography}{}
\expandafter\ifx\csname natexlab\endcsname\relax\def\natexlab#1{#1}\fi
\providecommand{\url}[1]{\href{#1}{#1}}
\providecommand{\dodoi}[1]{doi:~\href{http://doi.org/#1}{\nolinkurl{#1}}}
\providecommand{\doeprint}[1]{\href{http://ascl.net/#1}{\nolinkurl{http://ascl.net/#1}}}
\providecommand{\doarXiv}[1]{\href{https://arxiv.org/abs/#1}{\nolinkurl{https://arxiv.org/abs/#1}}}

\bibitem[{{Adams} {et~al.}(2024){Adams}, {Conselice}, {Austin}, {Harvey}, {Ferreira}, {Trussler}, {Juod{\v{z}}balis}, {Li}, {Windhorst}, {Cohen}, {Jansen}, {Summers}, {Tompkins}, {Driver}, {Robotham}, {D'Silva}, {Yan}, {Coe}, {Frye}, {Grogin}, {Koekemoer}, {Marshall}, {Pirzkal}, {Ryan}, {Maksym}, {Rutkowski}, {Willmer}, {Hammel}, {Nonino}, {Bhatawdekar}, {Wilkins}, {Bradley}, {Broadhurst}, {Cheng}, {Dole}, {Hathi}, \& {Zitrin}}]{Adams_2024}
{Adams}, N.~J., {Conselice}, C.~J., {Austin}, D., {et~al.} 2024, \apj, 965, 169, \dodoi{10.3847/1538-4357/ad2a7b}

\bibitem[{{Agarwal} {et~al.}(2019){Agarwal}, {Cullen}, {Khochfar}, {Ceverino}, \& {Klessen}}]{Agarwal_2019}
{Agarwal}, B., {Cullen}, F., {Khochfar}, S., {Ceverino}, D., \& {Klessen}, R.~S. 2019, \mnras, 488, 3268, \dodoi{10.1093/mnras/stz1347}

\bibitem[{{Ahn} {et~al.}(2009){Ahn}, {Shapiro}, {Iliev}, {Mellema}, \& {Pen}}]{Ahn_2009}
{Ahn}, K., {Shapiro}, P.~R., {Iliev}, I.~T., {Mellema}, G., \& {Pen}, U.-L. 2009, \apj, 695, 1430, \dodoi{10.1088/0004-637X/695/2/1430}

\bibitem[{{Asada} {et~al.}(2024){Asada}, {Sawicki}, {Abraham}, {Brada{\v{c}}}, {Brammer}, {Desprez}, {Estrada-Carpenter}, {Iyer}, {Martis}, {Matharu}, {Mowla}, {Muzzin}, {Noirot}, {Sarrouh}, {Strait}, {Willott}, \& {Harshan}}]{Asada_2024}
{Asada}, Y., {Sawicki}, M., {Abraham}, R., {et~al.} 2024, \mnras, 527, 11372, \dodoi{10.1093/mnras/stad3902}

\bibitem[{{Atek} {et~al.}(2023){Atek}, {Chisholm}, {Alavi}, {Berg}, {Bezanson}, {Blaizot}, {Bouwens}, {Chemerynska}, {Dessauges-Zavadsky}, {Endsley}, {Furtak}, {Kneib}, {Labbe}, {Leclercq}, {Marques-Chaves}, {Mason}, {Naidu}, {Natarajan}, {Oesch}, {Richard}, {Rosdahl}, {Saldana Lopez}, {Schaerer}, {Stark}, {Trebitsch}, \& {Volonteri}}]{Atek_2023}
{Atek}, H., {Chisholm}, J., {Alavi}, A., {et~al.} 2023, {JWST's GLIMPSE: gravitational lensing \& NIRCam imaging to probe early galaxy formation and sources of reionization}, JWST Proposal. Cycle 2, ID. \#3293

\bibitem[{{Bennett} \& {Sijacki}(2020)}]{Bennett_Sijacki_2020}
{Bennett}, J.~S., \& {Sijacki}, D. 2020, \mnras, 499, 597, \dodoi{10.1093/mnras/staa2835}

\bibitem[{{Berlind} \& {Weinberg}(2002)}]{Berlind_Weinberg_2002}
{Berlind}, A.~A., \& {Weinberg}, D.~H. 2002, \apj, 575, 587, \dodoi{10.1086/341469}

\bibitem[{{Blas} {et~al.}(2011){Blas}, {Lesgourgues}, \& {Tram}}]{Blas_2011_classy}
{Blas}, D., {Lesgourgues}, J., \& {Tram}, T. 2011, \jcap, 2011, 034, \dodoi{10.1088/1475-7516/2011/07/034}

\bibitem[{{Bouwens} {et~al.}(2023{\natexlab{a}}){Bouwens}, {Illingworth}, {Oesch}, {Stefanon}, {Naidu}, {van Leeuwen}, \& {Magee}}]{Bouwens_2023}
{Bouwens}, R., {Illingworth}, G., {Oesch}, P., {et~al.} 2023{\natexlab{a}}, \mnras, 523, 1009, \dodoi{10.1093/mnras/stad1014}

\bibitem[{{Bouwens} {et~al.}(2022){Bouwens}, {Illingworth}, {Ellis}, {Oesch}, \& {Stefanon}}]{Bouwens_2022}
{Bouwens}, R.~J., {Illingworth}, G., {Ellis}, R.~S., {Oesch}, P., \& {Stefanon}, M. 2022, \apj, 940, 55, \dodoi{10.3847/1538-4357/ac86d1}

\bibitem[{{Bouwens} {et~al.}(2021){Bouwens}, {Oesch}, {Stefanon}, {Illingworth}, {Labb{\'e}}, {Reddy}, {Atek}, {Montes}, {Naidu}, {Nanayakkara}, {Nelson}, \& {Wilkins}}]{Bouwens_2021}
{Bouwens}, R.~J., {Oesch}, P.~A., {Stefanon}, M., {et~al.} 2021, \aj, 162, 47, \dodoi{10.3847/1538-3881/abf83e}

\bibitem[{{Bouwens} {et~al.}(2023{\natexlab{b}}){Bouwens}, {Stefanon}, {Brammer}, {Oesch}, {Herard-Demanche}, {Illingworth}, {Matthee}, {Naidu}, {van Dokkum}, \& {van Leeuwen}}]{Bouwens_2023_UVLF}
{Bouwens}, R.~J., {Stefanon}, M., {Brammer}, G., {et~al.} 2023{\natexlab{b}}, \mnras, 523, 1036, \dodoi{10.1093/mnras/stad1145}

\bibitem[{{Bromm}(2013)}]{Bromm_2013}
{Bromm}, V. 2013, Reports on Progress in Physics, 76, 112901, \dodoi{10.1088/0034-4885/76/11/112901}

\bibitem[{{Bromm} {et~al.}(2009){Bromm}, {Yoshida}, {Hernquist}, \& {McKee}}]{Bromm_2009}
{Bromm}, V., {Yoshida}, N., {Hernquist}, L., \& {McKee}, C.~F. 2009, \nat, 459, 49, \dodoi{10.1038/nature07990}

\bibitem[{{Cai} {et~al.}(2025){Cai}, {Li}, {Cai}, {Wu}, {Yu}, {Dickinson}, {Sun}, {Fan}, {Wang}, {Cullen}, {Bian}, {Lin}, \& {Zou}}]{Cai_2025}
{Cai}, S., {Li}, M., {Cai}, Z., {et~al.} 2025, arXiv e-prints, arXiv:2507.17820, \dodoi{10.48550/arXiv.2507.17820}

\bibitem[{{Castellano} {et~al.}(2022){Castellano}, {Fontana}, {Treu}, {Santini}, {Merlin}, {Leethochawalit}, {Trenti}, {Vanzella}, {Mestric}, {Bonchi}, {Belfiori}, {Nonino}, {Paris}, {Polenta}, {Roberts-Borsani}, {Boyett}, {Brada{\v{c}}}, {Calabr{\`o}}, {Glazebrook}, {Grillo}, {Mascia}, {Mason}, {Mercurio}, {Morishita}, {Nanayakkara}, {Pentericci}, {Rosati}, {Vulcani}, {Wang}, \& {Yang}}]{Castellano_2022}
{Castellano}, M., {Fontana}, A., {Treu}, T., {et~al.} 2022, \apjl, 938, L15, \dodoi{10.3847/2041-8213/ac94d0}

\bibitem[{{Chakraborty} \& {Choudhury}(2025)}]{Chakraborty_Choudhury_2025}
{Chakraborty}, A., \& {Choudhury}, T.~R. 2025, arXiv e-prints, arXiv:2503.07590, \dodoi{10.48550/arXiv.2503.07590}

\bibitem[{{Chemerynska} {et~al.}(2024){Chemerynska}, {Atek}, {Furtak}, {Zitrin}, {Greene}, {Dayal}, {Weibel}, {Fujimoto}, {Kokorev}, {Goulding}, {Williams}, {Nanayakkara}, {Bezanson}, {Brammer}, {Cutler}, {Labbe}, {Leja}, {Pan}, {Price}, {van Dokkum}, {Wang}, {Weaver}, \& {Whitaker}}]{Chemerynska_2024}
{Chemerynska}, I., {Atek}, H., {Furtak}, L.~J., {et~al.} 2024, \mnras, 531, 2615, \dodoi{10.1093/mnras/stae1260}

\bibitem[{{Chon} {et~al.}(2024){Chon}, {Hosokawa}, {Omukai}, \& {Schneider}}]{Chon_2024}
{Chon}, S., {Hosokawa}, T., {Omukai}, K., \& {Schneider}, R. 2024, \mnras, 530, 2453, \dodoi{10.1093/mnras/stae1027}

\bibitem[{{Cruz} {et~al.}(2025){Cruz}, {Mu{\~n}oz}, {Sabti}, \& {Kamionkowski}}]{Cruz_2024}
{Cruz}, H. A.~G., {Mu{\~n}oz}, J.~B., {Sabti}, N., \& {Kamionkowski}, M. 2025, \prd, 111, 083503, \dodoi{10.1103/PhysRevD.111.083503}

\bibitem[{{Cueto} {et~al.}(2024){Cueto}, {Hutter}, {Dayal}, {Gottl{\"o}ber}, {Heintz}, {Mason}, {Trebitsch}, \& {Yepes}}]{Cueto_2024}
{Cueto}, E.~R., {Hutter}, A., {Dayal}, P., {et~al.} 2024, \aap, 686, A138, \dodoi{10.1051/0004-6361/202349017}

\bibitem[{{Dekel} {et~al.}(2023){Dekel}, {Sarkar}, {Birnboim}, {Mandelker}, \& {Li}}]{Dekel_2023}
{Dekel}, A., {Sarkar}, K.~C., {Birnboim}, Y., {Mandelker}, N., \& {Li}, Z. 2023, \mnras, 523, 3201, \dodoi{10.1093/mnras/stad1557}

\bibitem[{{Di Cesare} {et~al.}(2023){Di Cesare}, {Graziani}, {Schneider}, {Ginolfi}, {Venditti}, {Santini}, \& {Hunt}}]{DiCesare_2023}
{Di Cesare}, C., {Graziani}, L., {Schneider}, R., {et~al.} 2023, \mnras, 519, 4632, \dodoi{10.1093/mnras/stac3702}

\bibitem[{{Donnan} {et~al.}(2024){Donnan}, {McLure}, {Dunlop}, {McLeod}, {Magee}, {Arellano-C{\'o}rdova}, {Barrufet}, {Begley}, {Bowler}, {Carnall}, {Cullen}, {Ellis}, {Fontana}, {Illingworth}, {Grogin}, {Hamadouche}, {Koekemoer}, {Liu}, {Mason}, {Santini}, \& {Stanton}}]{Donnan_2024}
{Donnan}, C.~T., {McLure}, R.~J., {Dunlop}, J.~S., {et~al.} 2024, \mnras, 533, 3222, \dodoi{10.1093/mnras/stae2037}

\bibitem[{{Ferrara} {et~al.}(2023){Ferrara}, {Pallottini}, \& {Dayal}}]{Ferrara_2023}
{Ferrara}, A., {Pallottini}, A., \& {Dayal}, P. 2023, \mnras, 522, 3986, \dodoi{10.1093/mnras/stad1095}

\bibitem[{{Finkelstein} \& {Bagley}(2022)}]{Finkelstein_Bagley_2022}
{Finkelstein}, S.~L., \& {Bagley}, M.~B. 2022, \apj, 938, 25, \dodoi{10.3847/1538-4357/ac89eb}

\bibitem[{{Finkelstein} {et~al.}(2023){Finkelstein}, {Bagley}, {Ferguson}, {Wilkins}, {Kartaltepe}, {Papovich}, {Yung}, {Arrabal Haro}, {Behroozi}, {Dickinson}, {Kocevski}, {Koekemoer}, {Larson}, {Le Bail}, {Morales}, {P{\'e}rez-Gonz{\'a}lez}, {Burgarella}, {Dav{\'e}}, {Hirschmann}, {Somerville}, {Wuyts}, {Bromm}, {Casey}, {Fontana}, {Fujimoto}, {Gardner}, {Giavalisco}, {Grazian}, {Grogin}, {Hathi}, {Hutchison}, {Jha}, {Jogee}, {Kewley}, {Kirkpatrick}, {Long}, {Lotz}, {Pentericci}, {Pierel}, {Pirzkal}, {Ravindranath}, {Ryan}, {Trump}, {Yang}, {Bhatawdekar}, {Bisigello}, {Buat}, {Calabr{\`o}}, {Castellano}, {Cleri}, {Cooper}, {Croton}, {Daddi}, {Dekel}, {Elbaz}, {Franco}, {Gawiser}, {Holwerda}, {Huertas-Company}, {Jaskot}, {Leung}, {Lucas}, {Mobasher}, {Pandya}, {Tacchella}, {Weiner}, \& {Zavala}}]{Finkelstein_2023}
{Finkelstein}, S.~L., {Bagley}, M.~B., {Ferguson}, H.~C., {et~al.} 2023, \apjl, 946, L13, \dodoi{10.3847/2041-8213/acade4}

\bibitem[{{Finkelstein} {et~al.}(2024){Finkelstein}, {Leung}, {Bagley}, {Dickinson}, {Ferguson}, {Papovich}, {Akins}, {Arrabal Haro}, {Dav{\'e}}, {Dekel}, {Kartaltepe}, {Kocevski}, {Koekemoer}, {Pirzkal}, {Somerville}, {Yung}, {Amor{\'\i}n}, {Backhaus}, {Behroozi}, {Bisigello}, {Bromm}, {Casey}, {Ch{\'a}vez Ortiz}, {Cheng}, {Chworowsky}, {Cleri}, {Cooper}, {Davis}, {de la Vega}, {Elbaz}, {Franco}, {Fontana}, {Fujimoto}, {Giavalisco}, {Grogin}, {Holwerda}, {Huertas-Company}, {Hirschmann}, {Iyer}, {Jogee}, {Jung}, {Larson}, {Lucas}, {Mobasher}, {Morales}, {Morley}, {Mukherjee}, {P{\'e}rez-Gonz{\'a}lez}, {Ravindranath}, {Rodighiero}, {Rowland}, {Tacchella}, {Taylor}, {Trump}, \& {Wilkins}}]{Finkelstein_2024}
{Finkelstein}, S.~L., {Leung}, G. C.~K., {Bagley}, M.~B., {et~al.} 2024, \apjl, 969, L2, \dodoi{10.3847/2041-8213/ad4495}

\bibitem[{{Foreman-Mackey}(2016)}]{Foreman-Mackey_2016_corner}
{Foreman-Mackey}, D. 2016, The Journal of Open Source Software, 1, 24, \dodoi{10.21105/joss.00024}

\bibitem[{{Foreman-Mackey} {et~al.}(2013){Foreman-Mackey}, {Hogg}, {Lang}, \& {Goodman}}]{Foreman-Mackey_2013_emcee}
{Foreman-Mackey}, D., {Hogg}, D.~W., {Lang}, D., \& {Goodman}, J. 2013, \pasp, 125, 306, \dodoi{10.1086/670067}

\bibitem[{{Fujimoto} {et~al.}(2025{\natexlab{a}}){Fujimoto}, {Naidu}, {Chisholm}, {Atek}, {Endsley}, {Kokorev}, {Furtak}, {Pan}, {Liu}, {Bromm}, {Venditti}, {Visbal}, {Sarmento}, {Weibel}, {Oesch}, {Brammer}, {Schaerer}, {Adamo}, {Berg}, {Bezanson}, {Bouwens}, {Chemerynska}, {Claeyssens}, {Dessauges-Zavadsky}, {Frebel}, {Korber}, {Labbe}, {Marques-Chaves}, {Matthee}, {McQuinn}, {Mu{\~n}oz}, {Natarajan}, {Saldana-Lopez}, {Suess}, {Volonteri}, \& {Zitrin}}]{Fujimoto_2025}
{Fujimoto}, S., {Naidu}, R.~P., {Chisholm}, J., {et~al.} 2025{\natexlab{a}}, \apj, 989, 46, \dodoi{10.3847/1538-4357/ade9a1}

\bibitem[{{Fujimoto} {et~al.}(2025{\natexlab{b}}){Fujimoto}, {Coe}, {Abdurro'uf}, {Abraham}, {Adamo}, {Akins}, {Amorin}, {Arrabal Haro}, {Asada}, {Atek}, {Bagley}, {Bhatawdekar}, {Bradac}, {Bradley}, {Brammer}, {Bromm}, {Casey}, {Chisholm}, {Conselice}, {Dai}, {Dayal}, {Desprez}, {Dessauges-Zavadsky}, {Dickinson}, {Diego}, {Egami}, {Eisenstein}, {Ferguson}, {Finkelstein}, {Furtak}, {Hamilton}, {Harikane}, {Hashimoto}, {Hathi}, {Hsiao}, {Inayoshi}, {Jimenez-Teja}, {Jogee}, {Kartaltepe}, {Koekemoer}, {Kohno}, {Kokorev}, {Kumari}, {Labbe}, {Larson}, {Lucas}, {Magdis}, {Marchesini}, {Markov}, {Martis}, {Matthee}, {Meena}, {Messa}, {Mowla}, {Munoz}, {Naidu}, {Nakajima}, {Nakane}, {Noirot}, {Oesch}, {Oguri}, {Ono}, {Ouchi}, {Pan}, {Papovich}, {Pascale}, {Pierel}, {Postman}, {Resseguier}, {Rest}, {Richard}, {Ricotti}, {Rigby}, {Sawicki}, {Schneider}, {Shimasaku}, {Strolger}, {Sun}, {Toft}, {Tripodi}, {Trussler}, {Tsujita}, {Umeda}, {Valentino}, {Vanzella}, {Venditti}, {Watson}, {Weaver}, {Welch}, {Willott},
  {Windhorst}, {Xu}, {Yanagisawa}, {Zackrisson}, \& {Zitrin}}]{Fujimoto_2025_VENUS}
{Fujimoto}, S., {Coe}, D., {Abdurro'uf}, A., {et~al.} 2025{\natexlab{b}}, {Vast Exploration for Nascent, Unexplored Sources (VENUS)}, JWST Proposal. Cycle 4, ID. \#6882

\bibitem[{{Furlanetto} \& {Mirocha}(2022)}]{Furlanetto_Mirocha_2022}
{Furlanetto}, S.~R., \& {Mirocha}, J. 2022, \mnras, 511, 3895, \dodoi{10.1093/mnras/stac310}

\bibitem[{{Furlanetto} {et~al.}(2017){Furlanetto}, {Mirocha}, {Mebane}, \& {Sun}}]{Furlanetto_2017}
{Furlanetto}, S.~R., {Mirocha}, J., {Mebane}, R.~H., \& {Sun}, G. 2017, \mnras, 472, 1576, \dodoi{10.1093/mnras/stx2132}

\bibitem[{{Gehrels}(1986)}]{Gehrels_1986}
{Gehrels}, N. 1986, \apj, 303, 336, \dodoi{10.1086/164079}

\bibitem[{{Gelli} {et~al.}(2024){Gelli}, {Mason}, \& {Hayward}}]{Gelli_2024}
{Gelli}, V., {Mason}, C., \& {Hayward}, C.~C. 2024, \apj, 975, 192, \dodoi{10.3847/1538-4357/ad7b36}

\bibitem[{{Graziani} {et~al.}(2020){Graziani}, {Schneider}, {Ginolfi}, {Hunt}, {Maio}, {Glatzle}, \& {Ciardi}}]{Graziani_2020}
{Graziani}, L., {Schneider}, R., {Ginolfi}, M., {et~al.} 2020, \mnras, 494, 1071, \dodoi{10.1093/mnras/staa796}

\bibitem[{{Greif} \& {Bromm}(2006)}]{Greif_Bromm_2006}
{Greif}, T.~H., \& {Bromm}, V. 2006, \mnras, 373, 128, \dodoi{10.1111/j.1365-2966.2006.11017.x}

\bibitem[{{Greif} {et~al.}(2008){Greif}, {Johnson}, {Klessen}, \& {Bromm}}]{Greif_2008}
{Greif}, T.~H., {Johnson}, J.~L., {Klessen}, R.~S., \& {Bromm}, V. 2008, \mnras, 387, 1021, \dodoi{10.1111/j.1365-2966.2008.13326.x}

\bibitem[{{Haiman} {et~al.}(2000){Haiman}, {Abel}, \& {Rees}}]{Haiman_2000}
{Haiman}, Z., {Abel}, T., \& {Rees}, M.~J. 2000, \apj, 534, 11, \dodoi{10.1086/308723}

\bibitem[{{Harikane} {et~al.}(2024){Harikane}, {Nakajima}, {Ouchi}, {Umeda}, {Isobe}, {Ono}, {Xu}, \& {Zhang}}]{Harikane_2024}
{Harikane}, Y., {Nakajima}, K., {Ouchi}, M., {et~al.} 2024, \apj, 960, 56, \dodoi{10.3847/1538-4357/ad0b7e}

\bibitem[{{Harikane} {et~al.}(2023){Harikane}, {Ouchi}, {Oguri}, {Ono}, {Nakajima}, {Isobe}, {Umeda}, {Mawatari}, \& {Zhang}}]{Harikane_2023}
{Harikane}, Y., {Ouchi}, M., {Oguri}, M., {et~al.} 2023, \apjs, 265, 5, \dodoi{10.3847/1538-4365/acaaa9}

\bibitem[{{Harris} {et~al.}(2020){Harris}, {Millman}, {van der Walt}, {Gommers}, {Virtanen}, {Cournapeau}, {Wieser}, {Taylor}, {Berg}, {Smith}, {Kern}, {Picus}, {Hoyer}, {van Kerkwijk}, {Brett}, {Haldane}, {del R{\'\i}o}, {Wiebe}, {Peterson}, {G{\'e}rard-Marchant}, {Sheppard}, {Reddy}, {Weckesser}, {Abbasi}, {Gohlke}, \& {Oliphant}}]{Harris_2020_numpy}
{Harris}, C.~R., {Millman}, K.~J., {van der Walt}, S.~J., {et~al.} 2020, \nat, 585, 357, \dodoi{10.1038/s41586-020-2649-2}

\bibitem[{{Harvey} {et~al.}(2025){Harvey}, {Conselice}, {Adams}, {Austin}, {Juod{\v{z}}balis}, {Trussler}, {Li}, {Ormerod}, {Ferreira}, {Lovell}, {Duan}, {Westcott}, {Harris}, {Bhatawdekar}, {Coe}, {Cohen}, {Caruana}, {Cheng}, {Driver}, {Frye}, {Furtak}, {Grogin}, {Hathi}, {Holwerda}, {Jansen}, {Koekemoer}, {Marshall}, {Nonino}, {Vijayan}, {Wilkins}, {Windhorst}, {Willmer}, {Yan}, \& {Zitrin}}]{Harvey_2025}
{Harvey}, T., {Conselice}, C.~J., {Adams}, N.~J., {et~al.} 2025, \apj, 978, 89, \dodoi{10.3847/1538-4357/ad8c29}

\bibitem[{{Hirano} {et~al.}(2015){Hirano}, {Hosokawa}, {Yoshida}, {Omukai}, \& {Yorke}}]{Hirano_2015}
{Hirano}, S., {Hosokawa}, T., {Yoshida}, N., {Omukai}, K., \& {Yorke}, H.~W. 2015, \mnras, 448, 568, \dodoi{10.1093/mnras/stv044}

\bibitem[{{Hirano} {et~al.}(2014){Hirano}, {Hosokawa}, {Yoshida}, {Umeda}, {Omukai}, {Chiaki}, \& {Yorke}}]{Hirano_2014}
{Hirano}, S., {Hosokawa}, T., {Yoshida}, N., {et~al.} 2014, \apj, 781, 60, \dodoi{10.1088/0004-637X/781/2/60}

\bibitem[{{Hosokawa} {et~al.}(2016){Hosokawa}, {Hirano}, {Kuiper}, {Yorke}, {Omukai}, \& {Yoshida}}]{Hosokawa_2016}
{Hosokawa}, T., {Hirano}, S., {Kuiper}, R., {et~al.} 2016, \apj, 824, 119, \dodoi{10.3847/0004-637X/824/2/119}

\bibitem[{{Hosokawa} {et~al.}(2011){Hosokawa}, {Omukai}, {Yoshida}, \& {Yorke}}]{Hosokawa_2011}
{Hosokawa}, T., {Omukai}, K., {Yoshida}, N., \& {Yorke}, H.~W. 2011, Science, 334, 1250, \dodoi{10.1126/science.1207433}

\bibitem[{{Hunter}(2007)}]{Hunter_2007_matplotlib}
{Hunter}, J.~D. 2007, Computing in Science and Engineering, 9, 90, \dodoi{10.1109/MCSE.2007.55}

\bibitem[{{Hutter} {et~al.}(2025){Hutter}, {Cueto}, {Dayal}, {Gottl{\"o}ber}, {Trebitsch}, \& {Yepes}}]{Hutter_2025}
{Hutter}, A., {Cueto}, E.~R., {Dayal}, P., {et~al.} 2025, \aap, 694, A254, \dodoi{10.1051/0004-6361/202452460}

\bibitem[{{Inayoshi} {et~al.}(2022){Inayoshi}, {Harikane}, {Inoue}, {Li}, \& {Ho}}]{Inayoshi_2022}
{Inayoshi}, K., {Harikane}, Y., {Inoue}, A.~K., {Li}, W., \& {Ho}, L.~C. 2022, \apjl, 938, L10, \dodoi{10.3847/2041-8213/ac9310}

\bibitem[{{Ito} \& {Omukai}(2024)}]{Mana_Omukai_2024}
{Ito}, M., \& {Omukai}, K. 2024, \pasj, 76, 850, \dodoi{10.1093/pasj/psae054}

\bibitem[{{Jaacks} {et~al.}(2019){Jaacks}, {Finkelstein}, \& {Bromm}}]{Jaacks_2019}
{Jaacks}, J., {Finkelstein}, S.~L., \& {Bromm}, V. 2019, \mnras, 488, 2202, \dodoi{10.1093/mnras/stz1529}

\bibitem[{{Jeon} {et~al.}(2014){Jeon}, {Pawlik}, {Bromm}, \& {Milosavljevi{\'c}}}]{Jeon_2014}
{Jeon}, M., {Pawlik}, A.~H., {Bromm}, V., \& {Milosavljevi{\'c}}, M. 2014, \mnras, 444, 3288, \dodoi{10.1093/mnras/stu1980}

\bibitem[{{Jeong} {et~al.}(2025){Jeong}, {Jeon}, {Song}, \& {Bromm}}]{Jeong_2025}
{Jeong}, T.~B., {Jeon}, M., {Song}, H., \& {Bromm}, V. 2025, \apj, 980, 10, \dodoi{10.3847/1538-4357/ada27d}

\bibitem[{{Ji} {et~al.}(2015){Ji}, {Frebel}, \& {Bromm}}]{Ji_Mixing2015}
{Ji}, A.~P., {Frebel}, A., \& {Bromm}, V. 2015, \mnras, 454, 659, \dodoi{10.1093/mnras/stv2052}

\bibitem[{{Johnson} \& {Bromm}(2006)}]{Johnson_Bromm_2006}
{Johnson}, J.~L., \& {Bromm}, V. 2006, \mnras, 366, 247, \dodoi{10.1111/j.1365-2966.2005.09846.x}

\bibitem[{{Johnson} {et~al.}(2013){Johnson}, {Dalla Vecchia}, \& {Khochfar}}]{Johnson_2013}
{Johnson}, J.~L., {Dalla Vecchia}, C., \& {Khochfar}, S. 2013, \mnras, 428, 1857, \dodoi{10.1093/mnras/sts011}

\bibitem[{{Johnson} {et~al.}(2008){Johnson}, {Greif}, \& {Bromm}}]{Johnson_2008}
{Johnson}, J.~L., {Greif}, T.~H., \& {Bromm}, V. 2008, \mnras, 388, 26, \dodoi{10.1111/j.1365-2966.2008.13381.x}

\bibitem[{{Klessen} \& {Glover}(2023)}]{Klessen_Glover_2023}
{Klessen}, R.~S., \& {Glover}, S. C.~O. 2023, \araa, 61, 65, \dodoi{10.1146/annurev-astro-071221-053453}

\bibitem[{{Kokorev} {et~al.}(2025){Kokorev}, {Atek}, {Chisholm}, {Endsley}, {Chemerynska}, {Mu{\~n}oz}, {Furtak}, {Pan}, {Berg}, {Fujimoto}, {Oesch}, {Weibel}, {Adamo}, {Blaizot}, {Bouwens}, {Dessauges-Zavadsky}, {Khullar}, {Korber}, {Goovaerts}, {Jecmen}, {Labb{\'e}}, {Leclercq}, {Marques-Chaves}, {Mason}, {McQuinn}, {Naidu}, {Natarajan}, {Nelson}, {Rosdahl}, {Saldana-Lopez}, {Schaerer}, {Trebitsch}, {Volonteri}, \& {Zitrin}}]{Kokorev_2025}
{Kokorev}, V., {Atek}, H., {Chisholm}, J., {et~al.} 2025, \apjl, 983, L22, \dodoi{10.3847/2041-8213/adc458}

\bibitem[{{Latif} {et~al.}(2022){Latif}, {Whalen}, \& {Khochfar}}]{Latif_2022}
{Latif}, M.~A., {Whalen}, D., \& {Khochfar}, S. 2022, \apj, 925, 28, \dodoi{10.3847/1538-4357/ac3916}

\bibitem[{{Liu} \& {Bromm}(2020)}]{Liu_Bromm_2020}
{Liu}, B., \& {Bromm}, V. 2020, \mnras, 497, 2839, \dodoi{10.1093/mnras/staa2143}

\bibitem[{{Liu} {et~al.}(2025{\natexlab{a}}){Liu}, {Sibony}, {Meynet}, \& {Bromm}}]{Liu_2025}
{Liu}, B., {Sibony}, Y., {Meynet}, G., \& {Bromm}, V. 2025{\natexlab{a}}, \apjl, 980, L30, \dodoi{10.3847/2041-8213/adb151}

\bibitem[{{Liu} {et~al.}(2025{\natexlab{b}}){Liu}, {Kessler}, {Gessey-Jones}, {Dhandha}, {Fialkov}, {Sibony}, {Meynet}, {Bromm}, \& {Barkana}}]{Liu_2025_SAM}
{Liu}, B., {Kessler}, D., {Gessey-Jones}, T., {et~al.} 2025{\natexlab{b}}, \mnras, 541, 3113, \dodoi{10.1093/mnras/staf1178}

\bibitem[{{Lu} {et~al.}(2025){Lu}, {Frenk}, {Bose}, {Lacey}, {Cole}, {Baugh}, \& {Helly}}]{Lu_2025}
{Lu}, S., {Frenk}, C.~S., {Bose}, S., {et~al.} 2025, \mnras, 536, 1018, \dodoi{10.1093/mnras/stae2646}

\bibitem[{{Madau} \& {Dickinson}(2014)}]{Madau_Dickinson_2014}
{Madau}, P., \& {Dickinson}, M. 2014, \araa, 52, 415, \dodoi{10.1146/annurev-astro-081811-125615}

\bibitem[{{Magg} {et~al.}(2018){Magg}, {Hartwig}, {Agarwal}, {Frebel}, {Glover}, {Griffen}, \& {Klessen}}]{Magg_2018}
{Magg}, M., {Hartwig}, T., {Agarwal}, B., {et~al.} 2018, \mnras, 473, 5308, \dodoi{10.1093/mnras/stx2729}

\bibitem[{{Maiolino} {et~al.}(2024){Maiolino}, {{\"U}bler}, {Perna}, {Scholtz}, {D'Eugenio}, {Witten}, {Laporte}, {Witstok}, {Carniani}, {Tacchella}, {Baker}, {Arribas}, {Nakajima}, {Eisenstein}, {Bunker}, {Charlot}, {Cresci}, {Curti}, {Curtis-Lake}, {de Graaff}, {Egami}, {Ji}, {Johnson}, {Kumari}, {Looser}, {Maseda}, {Nelson}, {Robertson}, {Rodr{\'\i}guez Del Pino}, {Sandles}, {Simmonds}, {Smit}, {Sun}, {Venturi}, {Williams}, \& {Willmer}}]{Maiolino_2024}
{Maiolino}, R., {{\"U}bler}, H., {Perna}, M., {et~al.} 2024, \aap, 687, A67, \dodoi{10.1051/0004-6361/202347087}

\bibitem[{{Mason} {et~al.}(2023){Mason}, {Trenti}, \& {Treu}}]{Mason_2023}
{Mason}, C.~A., {Trenti}, M., \& {Treu}, T. 2023, \mnras, 521, 497, \dodoi{10.1093/mnras/stad035}

\bibitem[{{Mauerhofer} {et~al.}(2025){Mauerhofer}, {Dayal}, {Haehnelt}, {Kimm}, {Rosdahl}, \& {Teyssier}}]{Mauerhofer_2025}
{Mauerhofer}, V., {Dayal}, P., {Haehnelt}, M.~G., {et~al.} 2025, \aap, 696, A157, \dodoi{10.1051/0004-6361/202554042}

\bibitem[{{McLeod} {et~al.}(2024){McLeod}, {Donnan}, {McLure}, {Dunlop}, {Magee}, {Begley}, {Carnall}, {Cullen}, {Ellis}, {Hamadouche}, \& {Stanton}}]{McLeod_2024}
{McLeod}, D.~J., {Donnan}, C.~T., {McLure}, R.~J., {et~al.} 2024, \mnras, 527, 5004, \dodoi{10.1093/mnras/stad3471}

\bibitem[{{Mebane} {et~al.}(2018){Mebane}, {Mirocha}, \& {Furlanetto}}]{Mebane_2018}
{Mebane}, R.~H., {Mirocha}, J., \& {Furlanetto}, S.~R. 2018, \mnras, 479, 4544, \dodoi{10.1093/mnras/sty1833}

\bibitem[{{Menon} {et~al.}(2024){Menon}, {Lancaster}, {Burkhart}, {Somerville}, {Dekel}, \& {Krumholz}}]{Menon_2024}
{Menon}, S.~H., {Lancaster}, L., {Burkhart}, B., {et~al.} 2024, \apjl, 967, L28, \dodoi{10.3847/2041-8213/ad462d}

\bibitem[{{Mirocha} \& {Furlanetto}(2023)}]{Mirocha_Furlanetto_2023}
{Mirocha}, J., \& {Furlanetto}, S.~R. 2023, \mnras, 519, 843, \dodoi{10.1093/mnras/stac3578}

\bibitem[{{Morishita} {et~al.}(2025){Morishita}, {Liu}, {Stiavelli}, {Treu}, {Bergamini}, \& {Zhang}}]{Morishita_2025}
{Morishita}, T., {Liu}, Z., {Stiavelli}, M., {et~al.} 2025, arXiv e-prints, arXiv:2507.10521, \dodoi{10.48550/arXiv.2507.10521}

\bibitem[{{Moster} {et~al.}(2013){Moster}, {Naab}, \& {White}}]{Moster_2013}
{Moster}, B.~P., {Naab}, T., \& {White}, S. D.~M. 2013, \mnras, 428, 3121, \dodoi{10.1093/mnras/sts261}

\bibitem[{{Mu{\~n}oz}(2023)}]{Munoz_2023_Zeus21}
{Mu{\~n}oz}, J.~B. 2023, \mnras, 523, 2587, \dodoi{10.1093/mnras/stad1512}

\bibitem[{{Mu{\~n}oz} {et~al.}(2023){Mu{\~n}oz}, {Mirocha}, {Furlanetto}, \& {Sabti}}]{Munoz_2023}
{Mu{\~n}oz}, J.~B., {Mirocha}, J., {Furlanetto}, S., \& {Sabti}, N. 2023, \mnras, 526, L47, \dodoi{10.1093/mnrasl/slad115}

\bibitem[{{Mu{\~n}oz} {et~al.}(2022){Mu{\~n}oz}, {Qin}, {Mesinger}, {Murray}, {Greig}, \& {Mason}}]{Munoz_2022}
{Mu{\~n}oz}, J.~B., {Qin}, Y., {Mesinger}, A., {et~al.} 2022, \mnras, 511, 3657, \dodoi{10.1093/mnras/stac185}

\bibitem[{{Naidu} {et~al.}(2022){Naidu}, {Oesch}, {van Dokkum}, {Nelson}, {Suess}, {Brammer}, {Whitaker}, {Illingworth}, {Bouwens}, {Tacchella}, {Matthee}, {Allen}, {Bezanson}, {Conroy}, {Labbe}, {Leja}, {Leonova}, {Magee}, {Price}, {Setton}, {Strait}, {Stefanon}, {Toft}, {Weaver}, \& {Weibel}}]{Naidu_2022}
{Naidu}, R.~P., {Oesch}, P.~A., {van Dokkum}, P., {et~al.} 2022, \apjl, 940, L14, \dodoi{10.3847/2041-8213/ac9b22}

\bibitem[{{Nakajima} {et~al.}(2023){Nakajima}, {Ouchi}, {Isobe}, {Harikane}, {Zhang}, {Ono}, {Umeda}, \& {Oguri}}]{Nakajima_2023}
{Nakajima}, K., {Ouchi}, M., {Isobe}, Y., {et~al.} 2023, \apjs, 269, 33, \dodoi{10.3847/1538-4365/acd556}

\bibitem[{{Nikoli{\'c}} {et~al.}(2024){Nikoli{\'c}}, {Mesinger}, {Davies}, \& {Prelogovi{\'c}}}]{Nikolic_2024}
{Nikoli{\'c}}, I., {Mesinger}, A., {Davies}, J.~E., \& {Prelogovi{\'c}}, D. 2024, \aap, 692, A142, \dodoi{10.1051/0004-6361/202451213}

\bibitem[{{Oh} \& {Haiman}(2002)}]{Oh_Haiman_2002}
{Oh}, S.~P., \& {Haiman}, Z. 2002, \apj, 569, 558, \dodoi{10.1086/339393}

\bibitem[{{Oke}(1974)}]{Oke_1974}
{Oke}, J.~B. 1974, \apjs, 27, 21, \dodoi{10.1086/190287}

\bibitem[{{P{\'e}rez-Gonz{\'a}lez} {et~al.}(2023){P{\'e}rez-Gonz{\'a}lez}, {Costantin}, {Langeroodi}, {Rinaldi}, {Annunziatella}, {Ilbert}, {Colina}, {N{\o}rgaard-Nielsen}, {Greve}, {{\"O}stlin}, {Wright}, {Alonso-Herrero}, {{\'A}lvarez-M{\'a}rquez}, {Caputi}, {Eckart}, {Le F{\`e}vre}, {Labiano}, {Garc{\'\i}a-Mar{\'\i}n}, {Hjorth}, {Kendrew}, {Pye}, {Tikkanen}, {van der Werf}, {Walter}, {Ward}, {Bik}, {Boogaard}, {Bosman}, {G{\'o}mez}, {Gillman}, {Iani}, {Jermann}, {Melinder}, {Meyer}, {Moutard}, {van Dishoek}, {Henning}, {Lagage}, {Guedel}, {Peissker}, {Ray}, {Vandenbussche}, {Garc{\'\i}a-Argum{\'a}nez}, \& {Mar{\'\i}a M{\'e}rida}}]{Perez-Gonzalez_2023}
{P{\'e}rez-Gonz{\'a}lez}, P.~G., {Costantin}, L., {Langeroodi}, D., {et~al.} 2023, \apjl, 951, L1, \dodoi{10.3847/2041-8213/acd9d0}

\bibitem[{{Planck Collaboration}(2020)}]{Planck_2020}
{Planck Collaboration}. 2020, \aap, 641, A6, \dodoi{10.1051/0004-6361/201833910}

\bibitem[{{Qin} {et~al.}(2020){Qin}, {Mesinger}, {Park}, {Greig}, \& {Mu{\~n}oz}}]{Qin_2020}
{Qin}, Y., {Mesinger}, A., {Park}, J., {Greig}, B., \& {Mu{\~n}oz}, J.~B. 2020, \mnras, 495, 123, \dodoi{10.1093/mnras/staa1131}

\bibitem[{{Riaz} {et~al.}(2022){Riaz}, {Hartwig}, \& {Latif}}]{Riaz_2022}
{Riaz}, S., {Hartwig}, T., \& {Latif}, M.~A. 2022, \apjl, 937, L6, \dodoi{10.3847/2041-8213/ac8ea6}

\bibitem[{{Robertson} {et~al.}(2024){Robertson}, {Johnson}, {Tacchella}, {Eisenstein}, {Hainline}, {Arribas}, {Baker}, {Bunker}, {Carniani}, {Cargile}, {Carreira}, {Charlot}, {Chevallard}, {Curti}, {Curtis-Lake}, {D'Eugenio}, {Egami}, {Hausen}, {Helton}, {Jakobsen}, {Ji}, {Jones}, {Maiolino}, {Maseda}, {Nelson}, {P{\'e}rez-Gonz{\'a}lez}, {Pusk{\'a}s}, {Rieke}, {Smit}, {Sun}, {{\"U}bler}, {Whitler}, {Williams}, {Willmer}, {Willott}, \& {Witstok}}]{Robertson_2024}
{Robertson}, B., {Johnson}, B.~D., {Tacchella}, S., {et~al.} 2024, \apj, 970, 31, \dodoi{10.3847/1538-4357/ad463d}

\bibitem[{{Rodr{\'\i}guez-Puebla} {et~al.}(2016){Rodr{\'\i}guez-Puebla}, {Behroozi}, {Primack}, {Klypin}, {Lee}, \& {Hellinger}}]{Rodriguez-Puebla_2016}
{Rodr{\'\i}guez-Puebla}, A., {Behroozi}, P., {Primack}, J., {et~al.} 2016, \mnras, 462, 893, \dodoi{10.1093/mnras/stw1705}

\bibitem[{{Sabti} {et~al.}(2022){Sabti}, {Mu{\~n}oz}, \& {Blas}}]{Sabti_2022}
{Sabti}, N., {Mu{\~n}oz}, J.~B., \& {Blas}, D. 2022, \prd, 105, 043518, \dodoi{10.1103/PhysRevD.105.043518}

\bibitem[{{Sarmento} \& {Scannapieco}(2022)}]{Sarmento_Scannapieco_2022}
{Sarmento}, R., \& {Scannapieco}, E. 2022, \apj, 935, 174, \dodoi{10.3847/1538-4357/ac815c}

\bibitem[{{Sarmento} {et~al.}(2018){Sarmento}, {Scannapieco}, \& {Cohen}}]{Sarmento_2018}
{Sarmento}, R., {Scannapieco}, E., \& {Cohen}, S. 2018, \apj, 854, 75, \dodoi{10.3847/1538-4357/aa989a}

\bibitem[{{Schauer} {et~al.}(2022){Schauer}, {Bromm}, {Drory}, \& {Boylan-Kolchin}}]{Schauer_2022}
{Schauer}, A. T.~P., {Bromm}, V., {Drory}, N., \& {Boylan-Kolchin}, M. 2022, \apjl, 934, L6, \dodoi{10.3847/2041-8213/ac7f9a}

\bibitem[{{Shen} {et~al.}(2023){Shen}, {Vogelsberger}, {Boylan-Kolchin}, {Tacchella}, \& {Kannan}}]{Shen_2023}
{Shen}, X., {Vogelsberger}, M., {Boylan-Kolchin}, M., {Tacchella}, S., \& {Kannan}, R. 2023, \mnras, 525, 3254, \dodoi{10.1093/mnras/stad2508}

\bibitem[{{Sheth} \& {Tormen}(2002)}]{Sheth_Tormen_2002}
{Sheth}, R.~K., \& {Tormen}, G. 2002, \mnras, 329, 61, \dodoi{10.1046/j.1365-8711.2002.04950.x}

\bibitem[{{Shuntov} {et~al.}(2025){Shuntov}, {Oesch}, {Toft}, {Meyer}, {Covelo-Paz}, {Paquereau}, {Bouwens}, {Brammer}, {Gelli}, {Giovinazzo}, {Herard-Demanche}, {Illingworth}, {Mason}, {Naidu}, {Weibel}, \& {Xiao}}]{Shuntov_2025}
{Shuntov}, M., {Oesch}, P.~A., {Toft}, S., {et~al.} 2025, \aap, 699, A231, \dodoi{10.1051/0004-6361/202554618}

\bibitem[{{Skinner} \& {Wise}(2020)}]{Skinner_Wise_2020}
{Skinner}, D., \& {Wise}, J.~H. 2020, \mnras, 492, 4386, \dodoi{10.1093/mnras/staa139}

\bibitem[{{Somerville} {et~al.}(2025){Somerville}, {Yung}, {Lancaster}, {Menon}, {Sommovigo}, \& {Finkelstein}}]{Somerville_2025}
{Somerville}, R.~S., {Yung}, L.~Y.~A., {Lancaster}, L., {et~al.} 2025, arXiv e-prints, arXiv:2505.05442, \dodoi{10.48550/arXiv.2505.05442}

\bibitem[{{Stacy} {et~al.}(2016){Stacy}, {Bromm}, \& {Lee}}]{Stacy_2016}
{Stacy}, A., {Bromm}, V., \& {Lee}, A.~T. 2016, \mnras, 462, 1307, \dodoi{10.1093/mnras/stw1728}

\bibitem[{{Sugimura} {et~al.}(2020){Sugimura}, {Matsumoto}, {Hosokawa}, {Hirano}, \& {Omukai}}]{Sugimura_2020}
{Sugimura}, K., {Matsumoto}, T., {Hosokawa}, T., {Hirano}, S., \& {Omukai}, K. 2020, \apjl, 892, L14, \dodoi{10.3847/2041-8213/ab7d37}

\bibitem[{{Sun} {et~al.}(2023){Sun}, {Faucher-Gigu{\`e}re}, {Hayward}, {Shen}, {Wetzel}, \& {Cochrane}}]{Sun_2023}
{Sun}, G., {Faucher-Gigu{\`e}re}, C.-A., {Hayward}, C.~C., {et~al.} 2023, \apjl, 955, L35, \dodoi{10.3847/2041-8213/acf85a}

\bibitem[{{Susa} {et~al.}(2014){Susa}, {Hasegawa}, \& {Tominaga}}]{Susa_2014}
{Susa}, H., {Hasegawa}, K., \& {Tominaga}, N. 2014, \apj, 792, 32, \dodoi{10.1088/0004-637X/792/1/32}

\bibitem[{{Tang} \& {Chen}(2024)}]{Tang_Chen_2024}
{Tang}, C.-Y., \& {Chen}, K.-J. 2024, \mnras, 529, 4248, \dodoi{10.1093/mnras/stae764}

\bibitem[{{Taylor} {et~al.}(2025){Taylor}, {Kokorev}, {Kocevski}, {Akins}, {Cullen}, {Dickinson}, {Finkelstein}, {Arrabal Haro}, {Bromm}, {Giavalisco}, {Inayoshi}, {Juneau}, {Leung}, {P{\'e}rez-Gonz{\'a}lez}, {Somerville}, {Trump}, {Amor{\'\i}n}, {Barro}, {Burgarella}, {Brooks}, {Carnall}, {Casey}, {Cheng}, {Chisholm}, {Chworowsky}, {Davis}, {Donnan}, {Dunlop}, {Ellis}, {Fern{\'a}ndez}, {Fujimoto}, {Grogin}, {Gupta}, {Hathi}, {Jung}, {Hirschmann}, {Kartaltepe}, {Koekemoer}, {Larson}, {Leung}, {Llerena}, {Lucas}, {McLeod}, {McLure}, {Napolitano}, {Papovich}, {Stanton}, {Tripodi}, {Wang}, {Wilkins}, {Yung}, \& {Zavala}}]{Taylor_2024}
{Taylor}, A.~J., {Kokorev}, V., {Kocevski}, D.~D., {et~al.} 2025, \apjl, 989, L7, \dodoi{10.3847/2041-8213/ade789}

\bibitem[{{Trenti} \& {Stiavelli}(2008)}]{Trenti_Stiavelli_2008}
{Trenti}, M., \& {Stiavelli}, M. 2008, \apj, 676, 767, \dodoi{10.1086/528674}

\bibitem[{{Trinca} {et~al.}(2024){Trinca}, {Schneider}, {Valiante}, {Graziani}, {Ferrotti}, {Omukai}, \& {Chon}}]{Trinca_2024}
{Trinca}, A., {Schneider}, R., {Valiante}, R., {et~al.} 2024, \mnras, 529, 3563, \dodoi{10.1093/mnras/stae651}

\bibitem[{{van der Walt} {et~al.}(2011){van der Walt}, {Colbert}, \& {Varoquaux}}]{VanDerWalt_2011_numpy}
{van der Walt}, S., {Colbert}, S.~C., \& {Varoquaux}, G. 2011, Computing in Science and Engineering, 13, 22, \dodoi{10.1109/MCSE.2011.37}

\bibitem[{{Vanzella} {et~al.}(2020){Vanzella}, {Meneghetti}, {Caminha}, {Castellano}, {Calura}, {Rosati}, {Grillo}, {Dijkstra}, {Gronke}, {Sani}, {Mercurio}, {Tozzi}, {Nonino}, {Cristiani}, {Mignoli}, {Pentericci}, {Gilli}, {Treu}, {Caputi}, {Cupani}, {Fontana}, {Grazian}, \& {Balestra}}]{Vanzella_2020}
{Vanzella}, E., {Meneghetti}, M., {Caminha}, G.~B., {et~al.} 2020, \mnras, 494, L81, \dodoi{10.1093/mnrasl/slaa041}

\bibitem[{{Vanzella} {et~al.}(2023){Vanzella}, {Loiacono}, {Bergamini}, {Me{\v{s}}tri{\'c}}, {Castellano}, {Rosati}, {Meneghetti}, {Grillo}, {Calura}, {Mignoli}, {Brada{\v{c}}}, {Adamo}, {Rihtar{\v{s}}i{\v{c}}}, {Dickinson}, {Gronke}, {Zanella}, {Annibali}, {Willott}, {Messa}, {Sani}, {Acebron}, {Bolamperti}, {Comastri}, {Gilli}, {Caputi}, {Ricotti}, {Gruppioni}, {Ravindranath}, {Mercurio}, {Strait}, {Martis}, {Pascale}, {Caminha}, {Annunziatella}, \& {Nonino}}]{Vanzella_2023}
{Vanzella}, E., {Loiacono}, F., {Bergamini}, P., {et~al.} 2023, \aap, 678, A173, \dodoi{10.1051/0004-6361/202346981}

\bibitem[{{Venditti} {et~al.}(2024{\natexlab{a}}){Venditti}, {Bromm}, {Finkelstein}, {Calabr{\`o}}, {Napolitano}, {Graziani}, \& {Schneider}}]{Venditti_2024_HeII}
{Venditti}, A., {Bromm}, V., {Finkelstein}, S.~L., {et~al.} 2024{\natexlab{a}}, \apjl, 973, L12, \dodoi{10.3847/2041-8213/ad7387}

\bibitem[{{Venditti} {et~al.}(2024{\natexlab{b}}){Venditti}, {Bromm}, {Finkelstein}, {Graziani}, \& {Schneider}}]{Venditti_2024_PISNe}
{Venditti}, A., {Bromm}, V., {Finkelstein}, S.~L., {Graziani}, L., \& {Schneider}, R. 2024{\natexlab{b}}, \mnras, 527, 5102, \dodoi{10.1093/mnras/stad3513}

\bibitem[{{Venditti} {et~al.}(2023){Venditti}, {Graziani}, {Schneider}, {Pentericci}, {Di Cesare}, {Maio}, \& {Omukai}}]{Venditti_2023}
{Venditti}, A., {Graziani}, L., {Schneider}, R., {et~al.} 2023, \mnras, 522, 3809, \dodoi{10.1093/mnras/stad1201}

\bibitem[{Venditti {et~al.}(2025)Venditti, Munoz~Bermejo, Bromm, Fujimoto, Finkelstein, \& Chisholm}]{Venditti_2025_notebook}
Venditti, A., Munoz~Bermejo, J., Bromm, V., {et~al.} 2025, Bursty or heavy? Bright Population III systems in the Reionization era — modified Zeus21 with supporting data and notebook, v1,  Zenodo, \dodoi{10.5281/zenodo.16907335}

\bibitem[{{Ventura} {et~al.}(2024){Ventura}, {Qin}, {Balu}, \& {Wyithe}}]{Ventura_2024}
{Ventura}, E.~M., {Qin}, Y., {Balu}, S., \& {Wyithe}, J. S.~B. 2024, \mnras, 529, 628, \dodoi{10.1093/mnras/stae567}

\bibitem[{{Ventura} {et~al.}(2025){Ventura}, {Qin}, {Balu}, \& {Wyithe}}]{Ventura_2025}
---. 2025, \mnras, 540, 483, \dodoi{10.1093/mnras/staf699}

\bibitem[{{Virtanen} {et~al.}(2020){Virtanen}, {Gommers}, {Oliphant}, {Haberland}, {Reddy}, {Cournapeau}, {Burovski}, {Peterson}, {Weckesser}, {Bright}, {van der Walt}, {Brett}, {Wilson}, {Millman}, {Mayorov}, {Nelson}, {Jones}, {Kern}, {Larson}, {Carey}, {Polat}, {Feng}, {Moore}, {VanderPlas}, {Laxalde}, {Perktold}, {Cimrman}, {Henriksen}, {Quintero}, {Harris}, {Archibald}, {Ribeiro}, {Pedregosa}, {van Mulbregt}, \& {SciPy 1. 0 Contributors}}]{Virtanen_2020_scipy}
{Virtanen}, P., {Gommers}, R., {Oliphant}, T.~E., {et~al.} 2020, Nature Methods, 17, 261, \dodoi{10.1038/s41592-019-0686-2}

\bibitem[{{Visbal} {et~al.}(2020){Visbal}, {Bryan}, \& {Haiman}}]{Visbal_2020}
{Visbal}, E., {Bryan}, G.~L., \& {Haiman}, Z. 2020, \apj, 897, 95, \dodoi{10.3847/1538-4357/ab994e}

\bibitem[{{Wang} {et~al.}(2024){Wang}, {Cheng}, {Ge}, {Meng}, {Daddi}, {Yan}, {Ji}, {Jin}, {Jones}, {Malkan}, {Arrabal Haro}, {Brammer}, {Oguri}, {Hou}, \& {Zhang}}]{Wang_2024}
{Wang}, X., {Cheng}, C., {Ge}, J., {et~al.} 2024, \apjl, 967, L42, \dodoi{10.3847/2041-8213/ad4ced}

\bibitem[{{Welch} {et~al.}(2022){Welch}, {Coe}, {Diego}, {Zitrin}, {Zackrisson}, {Dimauro}, {Jim{\'e}nez-Teja}, {Kelly}, {Mahler}, {Oguri}, {Timmes}, {Windhorst}, {Florian}, {de Mink}, {Avila}, {Anderson}, {Bradley}, {Sharon}, {Vikaeus}, {McCandliss}, {Brada{\v{c}}}, {Rigby}, {Frye}, {Toft}, {Strait}, {Trenti}, {Sharma}, {Andrade-Santos}, \& {Broadhurst}}]{Welch_2022}
{Welch}, B., {Coe}, D., {Diego}, J.~M., {et~al.} 2022, \nat, 603, 815, \dodoi{10.1038/s41586-022-04449-y}

\bibitem[{{Wolcott-Green} \& {Haiman}(2019)}]{Wolcott-Green_2019}
{Wolcott-Green}, J., \& {Haiman}, Z. 2019, \mnras, 484, 2467, \dodoi{10.1093/mnras/sty3280}

\bibitem[{{Xu} {et~al.}(2016{\natexlab{a}}){Xu}, {Ahn}, {Norman}, {Wise}, \& {O'Shea}}]{Xu_2016_X-ray}
{Xu}, H., {Ahn}, K., {Norman}, M.~L., {Wise}, J.~H., \& {O'Shea}, B.~W. 2016{\natexlab{a}}, \apjl, 832, L5, \dodoi{10.3847/2041-8205/832/1/L5}

\bibitem[{{Xu} {et~al.}(2016{\natexlab{b}}){Xu}, {Norman}, {O'Shea}, \& {Wise}}]{Xu_2016}
{Xu}, H., {Norman}, M.~L., {O'Shea}, B.~W., \& {Wise}, J.~H. 2016{\natexlab{b}}, \apj, 823, 140, \dodoi{10.3847/0004-637X/823/2/140}

\bibitem[{{Yung} {et~al.}(2024){Yung}, {Somerville}, {Finkelstein}, {Wilkins}, \& {Gardner}}]{Yung_2024}
{Yung}, L.~Y.~A., {Somerville}, R.~S., {Finkelstein}, S.~L., {Wilkins}, S.~M., \& {Gardner}, J.~P. 2024, \mnras, 527, 5929, \dodoi{10.1093/mnras/stad3484}

\bibitem[{{Zier} {et~al.}(2025){Zier}, {Kannan}, {Smith}, {Puchwein}, {Vogelsberger}, {Borrow}, {Garaldi}, {Keating}, {McClymont}, {Shen}, \& {Hernquist}}]{Zier_2025}
{Zier}, O., {Kannan}, R., {Smith}, A., {et~al.} 2025, arXiv e-prints, arXiv:2503.03806, \dodoi{10.48550/arXiv.2503.03806}

\end{thebibliography}
\bibliographystyle{aasjournal}

%% This command is needed to show the entire author+affiliation list when
%% the collaboration and author truncation commands are used.  It has to
%% go at the end of the manuscript.
%\allauthors

%% Include this line if you are using the \added, \replaced, \deleted
%% commands to see a summary list of all changes at the end of the article.
%\listofchanges

\end{document}